\documentclass[12pt, draftclsnofoot, onecolumn]{IEEEtran}
\usepackage[tight,footnotesize]{subfigure}
\usepackage{float}
\usepackage{hyperref}
\usepackage{bm}
\usepackage{cite}
\usepackage{graphicx}
\usepackage[ruled]{algorithm2e} 
\usepackage{algorithmic}
\usepackage{amssymb}
\usepackage{amsmath}
\usepackage{amsthm}
\usepackage{mathrsfs}
\usepackage{color}
\usepackage{booktabs}

\theoremstyle{definition}
\newtheorem{theorem}{Theorem} 
\newtheorem{lemma}{Lemma}
\newtheorem{assumption}{Assumption}
\newtheorem{remark}{Remark}
\newtheorem{definition}{Definition}
\allowdisplaybreaks[1]
\allowdisplaybreaks

\newcommand{\tabincell}[2]{\begin{tabular}{@{}#1@{}}#2\end{tabular}} 

\begin{document}

\title{Knowledge-Guided Learning for  Transceiver Design in Over-the-Air Federated Learning}



\author{Yinan Zou, Zixin Wang, Xu Chen, Haibo Zhou, and Yong Zhou
	
	\thanks{Y. Zou, Z. Wang, and Y. Zhou are with the School of Information Science and Technology, ShanghaiTech University, Shanghai 201210, China (E-mail: \{zouyn, wangzx2, zhouyong\}@shanghaitech.edu.cn). 
	X. Chen is with the School of Computer Science and Engineering, Sun Yat-sen University, Guangzhou, China (e-mail: chenxu35@mail.sysu.edu.cn). 
	H. Zhou is with the School of Electronic Science and Engineering, Nanjing University, Nanjing 210023, China (e-mail: haibozhou@nju.edu.cn).} 
	}

\maketitle
\vspace{-15mm}
\begin{abstract}
In this paper, we consider communication-efficient over-the-air federated learning (FL), where multiple edge devices with non-independent and identically distributed datasets perform multiple local iterations in each communication round and then concurrently transmit their updated gradients to an edge server over the same radio channel for global model aggregation using over-the-air computation (AirComp). 
We derive the upper bound of the time-average norm of the gradients to characterize the convergence of AirComp-assisted FL, which reveals the impact of the model aggregation errors accumulated over all communication rounds on convergence. 
Based on the convergence analysis, we formulate an optimization problem to minimize the upper bound to enhance the learning performance, followed by proposing an alternating optimization algorithm to facilitate the optimal transceiver design for AirComp-assisted FL. 
As the alternating optimization algorithm suffers from high computation complexity, we further develop a knowledge-guided learning algorithm that exploits the structure of the analytic expression of the optimal transmit power to achieve computation-efficient transceiver design. 
Simulation results demonstrate that the proposed knowledge-guided learning algorithm achieves a comparable performance as the alternating optimization algorithm, but with a much lower computation complexity. 
Moreover, both proposed algorithms outperform the baseline methods in terms of  convergence speed and test accuracy.
\end{abstract}
\vspace{-5mm}
\begin{IEEEkeywords}
\vspace{-3mm}
Federated learning, over-the-air computation, knowledge-guided learning. 
\end{IEEEkeywords}

\section{Introduction}
With the ever increasing volume of distributed data, computing power of edge devices, and concerns of data privacy, federated  learning (FL) \cite{mcmahan2017communication,lim2020federated,lim2021dynamic} has recently been recognized as a promising distributed machine learning (ML) paradigm for edge artificial intelligence (AI) \cite{shi2020communication,shi2021mobile,rodrigues2019machine}. 
FL exploits the geographically dispersed data and computing power to distill intelligence at the network edge by employing an edge server to coordinate multiple edge devices to collaboratively train a shared ML model in an iterative manner. 
By executing the local training based on the local dataset and the up-to-date global model, each edge device only shares its model information instead of raw data  with the edge server to alleviate the privacy leakage concerns. 
FL is expected to support various intelligent applications \cite{yang2020FL}, including smart healthcare, industrial Internet of Things (IoT), and autonomous vehicles. 

FL over wireless networks has recently attracted considerable attention, where the communication-expensive model/gradient exchange between the edge server and edge devices is a critical issue that needs to be addressed. 
Because of limited radio spectrum resource and finite computing power of edge devices, it is crucial to study the communication and computation co-design. 
For instance, the authors in \cite{chen2020joint} proposed to jointly optimize the device selection, power control, and bandwidth allocation to minimize the FL training loss. 
By jointly optimizing the computation and communication resources, the authors in \cite{yang2020energy} developed an efficient algorithm to enable energy-efficient FL over wireless networks. 
Most existing studies adopt the orthogonal multiple access (OMA) scheme, e.g., frequency division multiple access (FDMA) and time division multiple access (TDMA), to ensure that the model update of each participating edge device is successfully received by the edge server before performing global model aggregation. 
Such a ``communicate-then-compute" strategy may not be spectrum-efficient as the number of required frequency/time resource blocks is proportional to  the number of participating edge devices.

Over-the-air computation (AirComp)\cite{nazer2007computation}, 
as an emerging non-orthogonal multiple access technique, has the potential to enable spectrum-efficient wireless model/gradient aggregation. 
By exploiting the waveform superposition nature, 
AirComp enables the edge server to receive a target  nomographic function (e.g., arithmetic mean, weighted average) of the signals concurrently transmitted by multiple  edge devices over the same radio channel.
During the model aggregation process of FL, the edge server is only interested in receiving a weighted average  of the local model updates from the edge devices, rather than each individual local model update. 
Such a model aggregation process matches well with the principle of AirComp, based on which the edge server can directly  obtain a noisy version of  the aggregated model update by allowing multiple edge devices to concurrently transmit their local model updates. 
Such a ``compute-when-communicate" strategy requires only one resource block regardless of the number of participating edge devices. 
The communication efficiency of AirComp-assisted FL has recently been demonstrated by the existing studies \cite{yang2020federated,zhu2019broadband,amiri2020machine} and further enhanced by leveraging intelligent reflecting surface (IRS) \cite{wang2021federated}. 

\subsection{Motivation and Contributions}
Most existing studies on AirComp-assisted FL \cite{yang2020federated, zhang2021gradient, wang2021federated, liu2021reconfigurable, xu2021learning} mainly treat each communication round equally important, and optimize the learning performance according to the instantaneous mean squared error (MSE) of the aggregated global model at a typical communication round, which leads to a sub-optimal learning performance \cite{xu2020client,zheng2020design}. 
This is because these studies ignore an inherent property of FL, i.e., the training process of FL involves multiple communication rounds and the model aggregation errors across all communication rounds collectively affect the final training performance. 
On the other hand, the existing studies \cite{yang2020federated, zhang2021gradient, wang2021federated, liu2021reconfigurable, xu2021learning, liu2020privacy, xiaowen2021optimized} mainly adopt optimization-based methods for the transceiver design of AirComp. 
However, the optimization-based methods typically suffer from high computation complexity and require the global channel state information (CSI),
which hinder their practical applications.
These two issues motivate us to develop a both communication and computation efficient 
framework to design, analyze, and optimize 
AirComp-assisted FL, taking into account the impact of the aggregation errors over all communication rounds on the FL performance.

In this paper, we consider over-the-air FL over a single-cell wireless network, where the edge devices with non-independent and identically distributed (non-i.i.d.)
datasets first perform multiple local iterations and  then concurrently transmit their gradients to the edge server over the same radio channel using AirComp in each communication round. 
Under this setup, we aim to characterize the convergence of the proposed communication-efficient AirComp-assisted FL and further develop a learning-based resource allocation algorithm to enhance the transceiver design, taking into account the model aggregation errors accumulated over all communication rounds. 
AirComp-assisted FL and learning-based transceiver design, as two critical components of our proposed unified framework, can be regarded as communication for AI and AI for communication, respectively. 
The main contributions of this paper are summarized as follows. 

\begin{itemize}
	\item 
	We theoretically analyze the convergence of the proposed communication-efficient  AirComp-assisted FL system, taking into account multiple local stochastic gradient descent (SGD) iterations and the non-i.i.d. data at edge devices.
	The convergence analysis demonstrates that the  time-average MSE is a critical  factor that captures the model aggregation errors accumulated over all communication rounds and determines the convergence performance of AirComp-assisted FL.
	To  enhance the learning performance, we formulate an optimization problem to minimize the time-average MSE of the aggregated global model, while taking into account the maximum and average transmit power budgets of each edge device.

	\item
	 To minimize the time-average MSE of the aggregated global model,
	 we propose an alternating optimization algorithm to optimize  the  transmit power of each edge device and the receive normalizing factor at the edge server.
	Due to the non-convexity arsing from the coupling between  the transmit power of edge devices and the receive normalizing factor,
	we decouple the optimization variables and transform the non-convex optimization  problem into two convex subproblems.
	We further derive  the optimal receive normalizing factor and the optimal 
	transmit power of edge devices  by leveraging KKT conditions.
	\item 
	As the proposed  optimization-based algorithm demands  relatively high computation complexity and requires global CSI,
	we further develop a novel knowledge-guided learning algorithm, which constructs a deep neural network (DNN) with domain knowledge  to map  the instantaneous CSI to the transmit power of edge devices and the receive normalizing factor.
	By exploiting  the structure of the analytical expression of the optimal transmit power,
	the proposed knowledge-guided learning algorithm reduces the searching space of the transmit power and in turn achieves a lower computation complexity than 
	 the conventional optimization-based algorithms.
	Moreover, as collecting the optimal solutions to
	the optimization problem as labels is generally time-consuming, we adopt unsupervised learning to train the DNN specifically developed for effective AirComp transceiver design.
	\item
	Simulation results demonstrate that the proposed
	alternating optimization algorithm and knowledge-guided learning algorithm achieve  faster convergence rates and  better learning performance than the baseline methods, including full power method, channel inversion method, and knowledge-free learning method.
	Moreover, the proposed knowledge-guided  learning algorithm can achieve a comparable learning  performance compared to the  proposed alternating optimization algorithm, 
	but with a much lower computation complexity.
\end{itemize}

\subsection{Related Works}
\subsubsection{OMA-based FL}
Various studies have recently been proposed to optimize 
resource allocation for FL over wireless networks \cite{dinh2020federated,wang2019adaptive,shi2020joint,wadu2021joint,zeng2021energy,ren2020accelerating}.
In particular, the authors in \cite{dinh2020federated} proposed an FL algorithm for the scenario with  non-i.i.d. data and developed an efficient resource allocation algorithm to improve the training performance. 
The authors in \cite{wang2019adaptive}
proposed  to adapt the frequency of global model  aggregation
to minimize the training loss.
With a fixed training time budget, a joint bandwidth allocation and scheduling policy was proposed in \cite{shi2020joint}.
By considering imperfect CSI, the authors in \cite{wadu2021joint} proposed a joint device scheduling and resource allocation algorithm to improve the training performance. 
Taking into account the CPU-GPU heterogeneous computing, the authors in \cite{zeng2021energy} designed a joint computation and communication resource allocation scheme to enhance the energy-efficiency of FL.
In addition to wirelss resource allocation, learning parameters (e.g., batch-size) can be further adjusted to enhance FL.
To accelerate the training process, the authors in \cite{ren2020accelerating} proposed a co-design of batch-size selection and communication resource allocation that can  adapt to time-varying wireless channels. 
Note that all  the aforementioned studies adopted the OMA scheme, which may not be spectrum-efficient for uplink model aggregation, especially when the number of edge devices is large. 

\subsubsection{AirComp-assisted FL}
Leveraging  AirComp to support  wireless  FL has recently been studied  from different perspectives. 
In particular, the authors in \cite{yang2020federated} proposed a joint design of device selection and receive beamforming to improve the learning performance of AirComp-assisted FL.     
To mitigate the aggregation error induced by AirComp, the authors in \cite{zhang2021gradient} developed an efficient transmit power control strategy. 
To alleviate the communication bottleneck of AirComp, the authors in \cite{wang2021federated, liu2021reconfigurable} leveraged  IRS to mitigate the magnitude misalignment during model/gradient aggregation.
The authors in \cite{liu2020privacy}
proposed to exploit  receiver
noise as a source of randomness to ensure differential privacy. 
Moreover, the local learning rates can be further optimized to enhance the learning performance based on the channel conditions in \cite{xu2021learning}. 
To reduce the implementation complexity,
the authors in \cite{paul2021accelerated} utilized momentum-based gradient to update the global model. 
However, most existing studies on  AirComp-assisted FL  did not take into account the model aggregation error accumulated over all communication rounds, which determines the final learning performance.

\subsubsection{Deep learning for resource allocation}
Due to the recent advancement of deep learning (DL), DNN
can be applied to reduce the computation complexity of  optimization-based resource allocation  algorithms  in wireless networks.
The authors in \cite{sun2018learning,liang2019towards,li2021multicell} proposed to train  DNNs  for interference management and sum rate maximization. 
The authors in \cite{shen2020graph}
utilized the graph neural network (GNN) for IRS configuration, beamformer design, and power control. 
However, these data-driven methods generally require a large amount of 
training samples and lack interpretability and predictability \cite{he2019model}.
These issues can be tackled by  model-driven  methods that construct neural networks based on domain knowledge. 
The authors in \cite{gao2018comnet} proposed a model-driven DNN to replace the conventional orthogonal frequency-division multiplexing (OFDM) receiver.
For joint activity detection and channel estimation,
the authors in \cite{shi2021algorithm,zou2021learning} proposed to unfold the numerical iterative methods as the recurrent neural network (RNN).
Furthermore, the authors in 
 \cite{xia2019deep} exploited 
the structure of the optimal solutions to design DNN for fast beamforming design.
%
Leveraging model-driven DL to achieve computation-efficient transceiver design  for AirComp-assisted FL is a critical issue, which, however, has not been explored.

\subsection{Organization and Notations}
The rest of this paper is organized as follows.
In Section II, we describe the system model of AirComp-assisted FL.
We present the convergence analysis and problem formulation in Section III.
In Section IV, we propose an alternating optimization algorithm.
In Section V, we develop a novel knowledge-guided learning algorithm.  
In Section VI, the simulation results are provided.
Finally, the paper is concluded in Section VII.
 
We use $\mathbb{R}^{n}$ to denote the real domain of dimension $n$.
Italic, boldface lower-case, and boldface upper-case letters are used to denote scalar, vector, and matrix, respectively.
We denote $(\cdot)^{\text{T}}$ as the transpose and $(\cdot)^{\text{H}}$ as Hermitian transpose.
$\mathbb{E}[\cdot]$ denotes the statistical expectation operator and $\|\cdot\|$ denotes the Euclidean norm.

\section{System Model}
\subsection{Federated Learning Model}

Consider FL over a single-cell wireless network, where an edge server co-located with a single-antenna base station coordinates $K$ single-antenna  edge devices to collaboratively  train a shared ML model, as shown in Fig. 1. 
We denote the index set of edge devices as $\mathcal{K} = \{1, \ldots, K  \}$. 
Each edge device $k\in\mathcal{K}$ owns a local dataset $\mathcal{D}_k = \{(\bm{x}_{k,i},y_{k,i}), 1\leq i \leq |\mathcal{D}_k| \}$, where $\bm{x}_{k,i}$ and $y_{k,i}$ denote the $i$-th data sample and its associated label at edge device $k$, respectively, and $|\mathcal{D}_k|$ denotes the cardinality of set $\mathcal{D}_k$.
The local data at the $k$-th edge device are generated according to the data distribution $\mathcal{T}_k, \forall \, k$.
In practice, the local data at different edge devices are usually non-i.i.d., i.e., $\mathcal{T}_k \neq \mathcal{T}_j, \forall \, k\neq j \in \mathcal{K}$. 
The local loss function at edge device $k$ with respect to local model vector  $\bm{w}_k \in \mathbb{R}^N$ of dimension $N$ is defined by the empirical risk over its local data
\vspace{-2mm}
\begin{align}\label{eq01}
	F_k(\bm{w}_k;\mathcal{D}_k)=\frac{1}{|\mathcal{D}_k|} \sum_{i=1}^{|\mathcal{D}_k|} f(\bm{w}_k;(\bm{x}_{k,i},y_{k,i})),
	\quad
	\forall \, k \in \mathcal{K},
\end{align}
where 
$f(\bm{w}_k;(\bm{x}_{k,i},y_{k,i}))$ denotes the sample-wise loss function with respect to $(\bm{x}_{k,i},y_{k,i})$.
For simplicity, we follow \cite{wang2021federated} and assume that each edge device has the same amount of data samples, i.e., $ |\mathcal{D}_1|=\cdots=|\mathcal{D}_K|=\frac{1}{K}|\mathcal{D}_{\text{tot}}|$, where $\mathcal{D}_{\text{tot}}=\cup_{k=1}^{K}\mathcal{D}_k$ denotes the global dataset. 
The empirical global loss function with respect to global model vector  $\bm{w}\in\mathbb{R}^N$ over the global dataset, denoted as $F(\bm{w};\mathcal{D}_{\text{tot}})$, is 
\vspace{-0mm}
\begin{align}\label{eq02}
	\!F(\bm{w};\mathcal{D}_{\text{tot}})
	=\!\frac{1}{|\mathcal{D}_{\text{tot}}|}\!\sum_{k=1}^K \!|\mathcal{D}_k|F_k(\bm{w}_k;\!\mathcal{D}_k)
	= \!\frac{1}{K}\!\sum_{k=1}^K \!F_k(\bm{w}_k;\mathcal{D}_k).
\end{align}
The objective of the training procedure is to find the optimal weight vector $\bm{w}^*$ that minimizes the global loss function $F(\bm{w};\mathcal{D}_{\text{tot}})$, i.e.,
\vspace{-0mm}
\begin{align}\label{eq03}
	\bm{w}^*=\arg \mathop{\min}_{\bm{w}\in
		\mathbb{R}^N} 
	F(\bm{w};\mathcal{D}_{\text{tot}}).
\end{align}
 \begin{figure}[!t]
 	\centering
 	\includegraphics[width=0.55\linewidth]{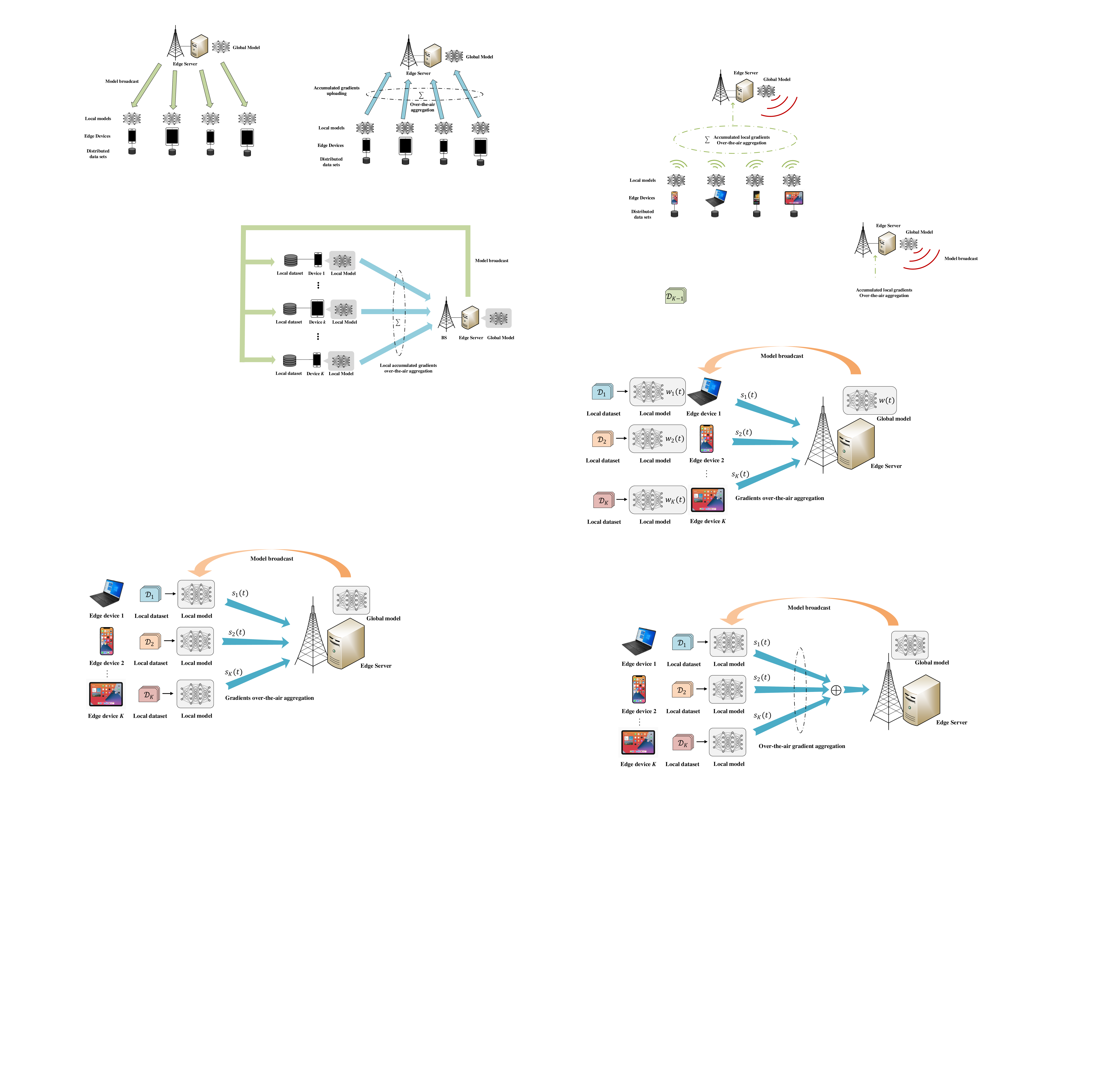}
 	\vspace{-0.4cm}
 	\caption{Illustration of AirComp-assisted FL.}
 	\label{systemmodel}
 	\vspace{-1cm}
 \end{figure}
\vspace{-0.8cm}
\subsection{Over-the-Air Federated Learning}
To achieve communication-efficient FL, we adopt the over-the-air  FedAvg algorithm, where all edge devices first execute multiple local iterations for local gradient computation and then concurrently transmit their  accumulated local gradients to the edge server using AirComp for global gradient aggregation. 
The computation and communication processes are elaborated as follows. 
At the beginning of communication round $t$, 
the edge server broadcasts global model $\bm{w}(t)$ to all edge devices in the downlink. 
As the edge server generally has a much higher transmit power than the edge devices, we assume that each edge device can receive  global model $\bm{w}(t)$ with negligible  distortion \cite{liu2020privacy}. 
After receiving global  model vector $\bm{w}(t)$, each edge device $k$ initializes its local model by setting $\bm{w}_k(t,0) = \bm{w}(t)$, and then updates its local model for $\phi > 1$ iterations  using the local stochastic gradient as follows
\vspace{-0mm}
\begin{align}\label{eq04}
	\bm{w}_k(t,\zeta+1) = \bm{w}_k(t,\zeta) - \lambda 
	\tilde{\bm{g}}_k(t,\zeta) ,
	\quad
	\zeta = 0,\ldots,\phi-1,
\end{align}
where $\lambda$ is the learning rate,  $\bm{w}_k(t,\zeta)$ is the local model at device $k$ in round $t$ after $\zeta$ local iterations, and
\vspace{-2mm}
\begin{align}\label{eq05}
	\tilde{\bm{g}}_k(t,\zeta) 
	&= \nabla F_k(\bm{w}_k(t,\zeta);\mathcal{B}_k(t,\zeta)) 
	=  \frac{1}{B}\sum_{(\bm{x_k},y_k)\in\mathcal{B}_k(t,\zeta)} \nabla f(\bm{w}_k(t,\zeta);(\bm{x_k},y_k)) 
\end{align}
is the stochastic gradient evaluated using the mini-batch $\mathcal{B}_k(t,\zeta)$ that contains $B$ randomly sampled  data samples from the local dataset $\mathcal{D}_k$. 

To update the global model, the edge server needs to obtain the aggregation of the  accumulated local gradients. 
Although the OMA scheme can be adopted for gradient uploading in the uplink, e.g., TDMA, the number of required resource blocks scales linearly with the number of edge devices. 
When there are a large number of edge devices participating in FL training, the incurred communication latency may be very large and becomes the main performance-limiting factor. 
To this end, we resort to using the AirComp technique to reduce the communication latency and thus enable fast gradient aggregation for FL. 
Specifically, during the uplink transmission process, all  edge devices concurrently  transmit their accumulated local gradients  to the edge server with appropriate pre-processing  over the same radio channel.
By exploiting the waveform superposition property of multiple-access channels, the server is capable of directly receiving  an aggregation of all  accumulated local gradients.
By enabling all edge devices to transmit concurrently,  the communication latency introduced by AirComp is independent of the number of edge devices, thereby achieving communication-efficient global gradient aggregation.

To facilitate transmit power control for edge devices, 
we normalize the $N$-dimensional  accumulated local gradients $\bm{\theta}_{k}(t) \in \mathbb{R}^N \triangleq (\bm{w}_k(t,0)-\bm{w}_k(t,\phi))/\lambda = \sum_{\zeta=0}^{\phi-1} \tilde{\bm{g}}_k(t,\zeta)$ before the uplink transmission.    
In particular, after computing the local model, device $k$ 
computes the mean $\bar{\theta}_{k}(t)$ and  variance $\pi_{k}^2(t)$ of $\bm{\theta}_{k}(t)$ as follows 
\vspace{-1mm}
\begin{align}
	&\bar{\theta}_{k}(t) = \frac{1}{N}\sum_{j=1}^{N}\theta_{k,j}(t),
	\
	\forall \, k \in \mathcal{K}, \label{eq06}
	\\& 
	\pi_{k}^2(t) = \frac{1}{N}\sum_{j=1}^{N}\bigg(\theta_{k,j}(t)-\bar{\theta}_{k}(t)\bigg)^2,
	\
	\forall \, k \in \mathcal{K}, \label{eq07}
\end{align}
where $\theta_{k,j}(t)$ denotes the $j$-th element of $\bm{\theta}_{k}(t)$.
By denoting $\bar{\theta}(t) = \frac{1}{K}\sum_{k\in\mathcal{K}}\bar{\theta}_{k}(t)$ and $\pi^2(t)=\frac{1}{K}\sum_{k\in\mathcal{K}}\pi^2_{k}(t)$, device $k$ normalizes $\bm{\theta}_{k}(t)$ as
\vspace{-0mm}
\begin{align}\label{eq08}
	\bm{s}_{k}(t) 
	=  \frac{\bm{\theta}_{k}(t)-\bar{\theta}(t)}{\pi(t)},\  
	\forall \, k \in \mathcal{K},
\end{align}
and transmits $\bm{s}_{k}(t)$ to the edge server over wireless fading channels.
Note that $\bm{s}_k(t)$ has zero mean and unit variance, i.e., $\mathbb{E}[\bm{s}_{k}(t)\bm{s}_{k}(t)^{\text{T}}]=\bm{I}_N$. 

We consider block-fading channels, i.e., the channel coefficients remain invariant within each communication round but vary independently from one round to another.
In the $t$-th communication round, we denote the complex-valued channel coefficient between edge device $k$ and the edge server as $h_{k}(t)$,
which is assumed to be known by edge device $k$, as in \cite{zhu2019broadband,amiri2020machine,wang2021federated}.
Before transmission, signal 
$\bm{s}_{k}(t)$ is multiplied by a  pre-processing factor $\psi_{k}(t)$ to compensate for the phase distortion due to channel fading.
In particular, we set $\psi_{k}(t) = \frac{\sqrt{p_{k}(t)}h_{k}^{\text{H}}(t)}{|h_{k}(t)|}$, where $p_{k}(t)\geq 0$ denotes the transmit power of device $k$ in the  $t$-th communication round.
We assume that all edge devices are synchronized, which can be achieved by either utilizing a reference-clock \cite{abari2015airshare} or adopting the timing advance technique \cite{mahmood2019time}. 
The received signal of dimension $N$ at the edge server can be expressed as
\vspace{-0mm}
\begin{align}\label{eq09}
	\bm{y}(t) = \sum_{k=1}^K h_{k}(t) \psi_{k}(t) \bm{s}_{k}(t) +\bm{n}(t),
\end{align}
where $\bm{n}(t)\sim\mathcal{N}(\sigma^2\bm{I}_N)$ denotes the additive white Gaussian noise (AWGN) vector.  
Upon receiving the signal, we apply the receive normalizing factor $\eta(t)$ at the edge server for signal amplitude alignment and noise suppression. 
Hence, we have
\vspace{-0mm}
\begin{align}\label{eq10}
	\hat{\bm{s}}(t)
	= \frac{\bm{y}(t)}{\sqrt{\eta(t)}} 
	= \sum_{k=1}^K \frac{\sqrt{p_{k}(t)}|h_{k}(t)|}{\sqrt{\eta(t)}} \bm{s}_{k}(t) +  \frac{\bm{n}(t)}{\sqrt{\eta(t)}}.
\end{align}
Note that $\hat{\bm{s}}(t)$ is an estimation of the target variable $\bm{s}(t) = \sum_{k=1}^K \bm{s}_{k}(t) $.
After de-normalization, we obtain
\vspace{-0mm}
\begin{align}\label{eq11}
	\hat{\bm{\theta}}(t)
	&= \frac{1}{K}\bigg(\pi(t) \hat{\bm{s}}(t) + K\bar{\theta}(t)\bigg).
\end{align}
Recall that $\bm{\theta}_{k}(t) = \pi(t) \bm{s}_{k}(t) 	+ \bar{\theta}(t)$ and $\bm{s}(t) = \sum_{k=1}^K \bm{s}_{k}(t)$, (\ref{eq11}) can be rewritten as
\vspace{-0mm}
\begin{align}\label{eq12}
	\hat{\bm{\theta}}(t) = \frac{1}{K} \pi(t) \bigg(\hat{\bm{s}}(t)-\bm{s}(t)\bigg)
	+ \bm{\theta}(t),
\end{align}
where $\bm{\theta}(t) = \frac{1}{K} \sum_{k\in\mathcal{K}} \bm{\theta}_{k}(t)$.
The edge server can only obtain an estimation of $\bm{\theta}(t)$, i.e.,  
$\hat{\bm{\theta}}(t)$, to update the global model parameter $\bm{w}(t+1)$ as follows
\vspace{-0mm}
\begin{align}\label{eq13}
	\bm{w}(t+1) 
	= \bm{w}(t) - \lambda \hat{\bm{\theta}}(t)
	= \bm{w}(t) - \lambda (\bm{\theta}(t) + \bm{e}(t)),
\end{align}
where $\bm{e}(t)=\frac{1}{K} \pi(t) (\hat{\bm{s}}(t)-\bm{s}(t))$ represents the random aggregation error in each communication round. 
This error is introduced by channel fading and receiver noise, and determines the convergence performance of FL. 

The  learning process proceeds by performing (\ref{eq04}), (\ref{eq10}), and (\ref{eq13}) iteratively, until the global model is converged or the maximum number of communication rounds is reached.
In addition, we consider that each edge device has the following maximum transmit power constraint and average transmit power constraint
\vspace{-1mm}
\begin{align}
	&  p_{k}(t) 
	\leq P_k^{\text{max}}, 
	 \quad \forall \, k\in
	 \mathcal{K},\ \forall \, t\in\{0,\ldots,T-1\},  \label{eq14} 
	\\& 
	\frac{1}{T} \sum_{t=0}^{T-1} p_{k}(t)
	\leq \bar{P}_k,  
	\quad \forall \, k\in
	\mathcal{K}, \label{eq15}
\end{align}
where $P_k^{\text{max}}>0$ and $\bar{P}_k>0$ denote the maximum and average transmit power budgets of edge device $k$, respectively, and $T$ is the maximum number of communication rounds \cite{xiaowen2021optimized,fu2021uav,xu2020client}. 
To make the average transmit power constraint non-trival, we assume $\bar{P}_k < P_k^{\text{max}}$.

\section{Convergence Analysis and Problem Formulation}
In this section, we present the convergence analysis for AirComp-assisted FL, taking into account multiple local SGD iterations  and the non-i.i.d. data, followed by formulating an optimization problem to minimize the upper bound of the time-average norm of the gradients. 

\subsection{Preliminary}

\subsubsection{Non-i.i.d. data}
With non-i.i.d. datasets among edge devices, the local optimum of the local loss function may not be consistent with the global optimum of the global loss function. 
As the heterogeneity level of the local gradients reflects that of the local data, we define the following metric.

\vspace{-3mm}
\begin{definition}\label{definition1}
	For $K$ edge devices with local gradients $\{\nabla F_k(\bm{w}_k)\}$, 
	we define metric $\chi$ to characterize the heterogeneity level of the local gradients as follows
	\begin{align} \label{eq16}
		&\frac{\frac{1}{K}\sum_{k=1}^K\| \nabla F_k(\bm{w}_k) \|_2^2}{\| \frac{1}{K}\sum_{k=1}^K \nabla F_k(\bm{w}_k) \|_2^2} 
		=  
		\frac{\frac{1}{K}\sum_{k=1}^K\| \nabla F_k(\bm{w}_k) \|_2^2}
		{\!\frac{1}{K^2}\!\sum_{k=1}^K\!\|\nabla F_k(\bm{w}_k)\|_2^2+\!\frac{1}{K^2}\!\sum_{i\neq j}\!\langle \nabla F_i(\bm{w}_i),\nabla F_j(\bm{w}_j)\!\rangle} 
		\leq  \chi.
	\end{align}   
\end{definition}
\begin{remark}
   	The inner product between two local gradients indicates the divergence of the directions of these two local gradients.
   	Note that $\chi\geq 1$ due to Jensen's inequality.
	When the data across edge devices are i.i.d.,
	the local gradients tend to be the same with tremendous data samples, and thus $\chi=1$.
	However, with statistically   heterogeneous data,
	the data distributions among edge devices are different,
	which implies that the local gradients pointing to different directions.
	Hence,
	the inner product between the local gradients is small,
	leading to a large value of $\chi$.
	In particular, a higher level of non-i.i.d. data incurs a larger value of $\chi$.
\end{remark}

\subsubsection{Basic assumptions}
To facilitate the convergence analysis, we make the following assumptions on the loss function and gradients. 

\vspace{-5mm}
\begin{assumption}[\textbf{Bounded loss function}]\label{assumption1}
	For any parameter $\bm{w}$, the global loss function is lower bounded, i.e., $F(\bm{w})\geq F(\bm{w}^*)>-\infty$.
\end{assumption}

\vspace{-8mm}
\begin{assumption}[\textbf{Lipshchitz continuity and smoothness}]\label{assumption2}
	The local loss function $F_k(\bm{w})$ is smooth with a non-negative constant $L$ and   continuously differentiable, i.e., 
	\vspace{-1mm}
	\begin{align}\label{eq17}
		\| \nabla F_k(\bm{w}) - \nabla F_k(\bm{w}')\|_2
		\leq L\| \bm{w} - \bm{w}'\|_2,
		\quad 
		\forall \, \bm{w},\bm{w}'.
	\end{align} 
	Inequality (\ref{eq17}) directly leads to  the following inequality 
	\vspace{-1mm}
	\begin{align}\label{eq18}
		F_k(\bm{w}') \leq F_k(\bm{w}) + \langle \nabla F_k(\bm{w}), \bm{w}'-\bm{w} \rangle + \frac{L}{2}\|\bm{w}-\bm{w}'\|^2_2,
		\quad \forall \, \bm{w}, \bm{w}'.
	\end{align}
\end{assumption}
\vspace{-6mm}
\begin{assumption}[\textbf{Bounded stochastic gradient variance}]\label{assumption3}
	The local mini-batch stochastic gradients $\{\tilde{\bm{g}}_k\}$ are assumed to be independent and unbiased estimates of the batch gradient $\{ \nabla F_k(\bm{w}_k) \}$ with bounded variance, i.e.,
	\vspace{-1mm}
	\begin{align}\label{eq19}
		\mathbb{E}[\tilde{\bm{g}}_k] = \nabla F_k(\bm{w}_k), \quad \forall \, k\in\mathcal{K},
	\end{align} 
	\begin{align}\label{eq20}
		\text{Var}(\tilde{\bm{g}}_k) = \mathbb{E}[\| \tilde{\bm{g}}_k - \nabla F_k(\bm{w}_k) \|_2^2] \leq \xi^2, \quad \forall \, k\in\mathcal{K},
	\end{align}
	where $\xi \geq 0 $ is a constant introduced to quantify the sampling noise of stochastic gradients.
\end{assumption}
\vspace{-8mm}
\begin{assumption} [\textbf{Bounded variance}]\label{assumption4}
	The variance of $N$ elements of $\bm{\theta}_k$ is upper bounded by a constant $\Gamma\geq0$, i.e., $\pi_k^2\leq\Gamma$. 
\end{assumption}
\vspace{-8mm}
\begin{remark}
	While Assumption \ref{assumption1} is necessary for converging to a stationary point \cite{bernstein2018signsgd}, Assumption \ref{assumption2} is standard for convergence analysis \cite{liu2020privacy}.
	Assumption \ref{assumption3} indicates that the stochastic gradient is an unbiased estimate of the batch gradient.
	Due to non-i.i.d. data  across edge devices, the local stochastic gradient is no longer an  unbiased estimate of the batch gradient of  the global loss function \cite{haddadpour2021federated}.
	For Assumption \ref{assumption4}, since the elements of $\bm{\theta}_{k}$ have finite values, 
	it is reasonable to assume that $\pi_k^2$, as the variance of $N$ elements of $\bm{\theta}_{k}$, is upper bounded as in \cite{wang2021federated}.
\end{remark}

\subsection{Convergence Analysis}
Based on the above assumptions, we present the following theorem for the convergence of AirComp-assisted FL with multiple local iterations and non-i.i.d. data.
\begin{theorem} \label{theorem}
	With Assumptions 1-4,
	if the learning rate $\lambda$ and the number of local iterations $\phi$ satisfy 
	$\phi^2L^2\lambda^2\chi+2\phi\lambda L \leq 1$, then the time-average norm of the gradients after $T$ communication rounds is upper bounded by 
	\begin{align}\label{eq21}
		\frac{1}{T}\sum_{t=0}^{T-1}\|\nabla F(\bm{w}(t))\|_2^2
		&\leq \underbrace{\frac{2(F(\bm{w}(0)-F(\bm{w}^*)))}{\lambda(\phi-1)T}}_{\text{Initial optimality gap}} 
		+ 
		\underbrace{\frac{2}{\phi-1}
			\left(\frac{\phi^2\lambda^2L^2}{2}+\frac{\phi\lambda L}{K}\right)\xi^2 }_{\text{Variance of stochatic gradient}}
		\notag
		\\& 
		+ 
		\underbrace{\frac{1+2\lambda L}{\phi-1}
		\frac{N\Gamma(K\!+\!1)}{K^2} \frac{1}{T}\sum_{t=0}^{T-1} \underbrace{ \text{MSE}(t)}_{\text{instantaneous MSE}}}_{\text{Time-average MSE}},
	\end{align}
	where 
	\vspace{-3mm}
	\begin{align}\label{eq22}
		\text{MSE}(t) =   \sum_{k=1}^K 
		\bigg( \frac{\sqrt{p_{k}(t)}|h_{k}(t)|}{\sqrt{\eta(t)}} - 1 \bigg)^2 
		+  \frac{\sigma^2}{\eta(t)}.
	\end{align}
\end{theorem}

\proof  Please refer to Appendix \ref{proof1}.  \qedhere
\vspace{-4mm}
\begin{remark} 
		For non-i.i.d. data, when the level of  the local gradient heterogeneity $\chi$ has a larger value, we set a smaller learning rate and perform a smaller number of local iterations to ensure that condition $\phi^2L^2\lambda^2\chi+2\phi\lambda L\leq1$ is satisfied.
\end{remark}
\vspace{-5mm}
\begin{remark}
		In Theorem \ref{theorem}, we adopt the average norm of the gradients  as the convergence indicator, which is widely used in the convergence analysis for non-convex loss \cite{wang2018cooperative}.
		Note that the FL algorithm achieves an $\epsilon$-approximation solution if 
		\begin{align}\label{eq23}
			\frac{1}{T}\sum_{t=0}^{T-1}\|\nabla F(\bm{w}(t)) \|_2^2 \leq \epsilon. 
		\end{align}
		We observe that the upper bound (\ref{eq21}) is composed of three terms.
		The first two terms are the initial optimality gap and  the variance of stochastic gradient.
		The last term is the time-average MSE resulting from analog gradient transmission.   
		As $T \to \infty$, the initial optimality gap decreases to zero, and the upper bound approaches to the summation of the variance of the stochastic gradient  and the time-average MSE.
		Besides, when the number of edge devices, the number of local iterations, and the learning rate are given, the variance of the stochastic gradient is a constant. 
		Consequently, in order to improve the convergence performance, it is necessary to minimize the time-average MSE given in \eqref{eq21}, which incorporates the model aggregation errors accumulated over $T$ communication rounds. 
\end{remark}

\vspace{-7mm}
\subsection{Problem Formulation}
By omitting the constant terms, we rewrite the time-average MSE as 
\begin{align} \label{eq24}
	\overline{\text{MSE}} 
	&= 
	\sum_{t=0}^{T-1} 
	\bigg[\sum_{k=1}^K 
	\bigg( \frac{\sqrt{p_{k}(t)}|h_{k}(t)|}{\sqrt{\eta(t)}} - 1 \bigg)^2 
	+  \frac{\sigma^2}{\eta(t)}\bigg]. 
\end{align} 
We aim to minimize the time-average MSE by jointly optimizing the transmit power $\{p_{k}(t)\}$ of edge devices and the receive normalizing factors $\{\eta(t)\}$ of the edge server.
Hence, the optimization problem can be formulated as 
\vspace{-1mm}
\begin{subequations} \label{eq25}
	\begin{align}  
		\mathscr{P}:
		\mathop{ \min} \limits_{ \{p_{k}(t)\},\atop \{\eta(t)\}   } \quad
		&\overline{\text{MSE}}
		\label{eq25a} \\
		\text{s.t.} \quad &0\leq  p_{k}(t)  \leq P_k^{\text{max}},
		\quad \forall \, k , \forall \, t, \label{eq25b}\\
		& 0\leq \frac{1}{T} \sum_{t=0}^{T-1}  p_{k}(t)
		\leq  \bar{P}_{k} , \quad \forall \, k, \label{eq25c}
		\\
		  &\eta(t) \geq 0, \quad  \forall \, t . 
	\end{align}
\end{subequations}

The objective function of problem $\mathscr{P}$ contains 
the noise-induced error (i.e., $\sum_{t=0}^{T-1}[\sigma^2/\eta(t)]$) and the signal misalignment error (i.e., $\sum_{t=0}^{T-1} 
[\sum_{k=1}^K 
( \sqrt{p_{k}(t)}|h_{k}(t)|/\sqrt{\eta(t)} - 1 )^2 
]$). 
To minimize the time-average MSE, an intuitive idea is to enlarge the receive normalizing factors  to diminish the noise-induced error and adjust the transmit power of edge devices to align the signal amplitudes. 
However, the finite average and maximum transmit power budgets make the signal amplitude alignment not always possible. 
Hence, it is tricky to simultaneously reduce the signal misalignment error and the noise-induced error.
Moreover, problem (\ref{eq25}) is a non-convex optimization problem as the transmit power of edge devices and the receive normalizing factor are highly coupled over different communication rounds. 
All these issues make problem $\mathscr{P}$ challenging to be solved. 

\section{Alternating Optimization Algorithm}\label{sec_opt}
In this section, we propose an alternating optimization algorithm to decouple the optimization variables and tackle the non-convex optimization problem $\mathscr{P}$.
\subsection{Receive Normalizing Factor Optimization}
We first optimize the receive normalizing factors $\{\eta(t)\}$ with given transmit power of  edge devices $\{p_{k}(t)\}$ by solving the following problem
\begin{align}\label{eq26}
	\mathscr{P}_{1}:\mathop{ \min} \limits_{\{\eta(t)\geq0\}  } 
	\!\sum_{t=0}^{T-1}\!  
	\bigg[ \!  \sum_{k=1}^K  \!
	\bigg( \! \frac{\sqrt{p_{k}(t)}|h_{k}(t)|}{\sqrt{\eta(t)}}\! - 1\! \bigg)^2 \!
	+  \frac{\sigma^2}{\eta(t)}  \bigg].
\end{align}
We decompose problem $\mathscr{P}_{1}$ into $T$ subproblems for $T$ communication rounds.
Each subproblem can be expressed as 
\vspace{-1mm}
\begin{align}\label{eq27}
	\mathop{ \min} \limits_{\eta(t)\geq0  } \mathcal{E}(\eta(t)) \triangleq \sum_{k=1}^K   \left(\frac{\sqrt{p_{k}(t)}|h_{k}(t)|}{\sqrt{\eta(t)}}-1\right)^2  
	+ \frac{\sigma^2}{\eta(t)},
\end{align}
where $\mathcal{E}(\eta(t))$ denotes the objective function of problem (\ref{eq27}). 
By denoting $\Omega(t)=1/\sqrt{\eta(t)}$, we rewrite the objective function of problem (\ref{eq27}) as
\begin{align}\label{eq28}
	\mathcal{E}(\Omega(t)) = \sum_{k=1}^K   \left(\sqrt{p_{k}(t)}|h_{k}(t)|\Omega(t)-1\right)^2  
	+ (\sigma\Omega(t))^2,
\end{align}
which is convex with respect to $\Omega(t)$.
By setting the first-order derivative of the objective function $\mathcal{E}(\Omega(t))$ to zero, we 
obtain the closed-form expression of the optimal $\Omega^{*}(t)$.
As a result, the optimal receive normalizing factor $\eta^{*}(t)$ to problem (\ref{eq27}) can be expressed as
\begin{align}\label{eq29}
\hspace{-2mm}	\eta^{*}(t) = \frac{1}{(\Omega^{*}(t))^2}
	= \bigg(\frac{\sigma^2+\sum_{k=1}^K(\sqrt{p_{k}(t)}|h_{k}(t)|)^2}{\sum_{k=1}^K\sqrt{p_{k}(t)}|h_{k}(t)|}\bigg)^2.
\end{align}

\subsection{Transmit Power Optimization}
We fix the obtained optimal receive normalizing factor and optimize $\{p_{k}(t)\}$  by solving the following problem
\vspace{-1mm}
\begin{subequations}\label{eq30}
\begin{align} 
\mathscr{P}_{2}:\mathop{ \min} \limits_{\{p_{k}(t)\}  }  
\quad&\sum_{t=0}^{T-1}\bigg[ \sum_{k=1}^K   \bigg(\frac{\sqrt{p_{k}(t)}|h_{k}(t)|}{\sqrt{\eta(t)}}-1\bigg)^2  
\bigg]
\\
\text{s.t.} \quad&  \text{constraints}\ \eqref{eq25b}\eqref{eq25c}.
\end{align}
\end{subequations}
We decompose problem $\mathscr{P}_{2}$ into $K$ subproblems and optimize the transmit power of the $k$-th edge device by solving the following problem
\begin{subequations}\label{eq31}
	\begin{align} 
		\mathop{ \min} \limits_{\{p_{k}(t)\}  } &\quad
		\sum_{t=0}^{T-1}   \left(\frac{\sqrt{p_{k}(t)}|h_{k}(t)|}{\sqrt{\eta(t)}}-1\right)^2  \label{eq31a}
		\\ 
		\text{s.t.}& \quad 0\leq  p_{k}(t)\leq P_k^{\text{max}},
		\quad   \forall \, t, \label{eq31b}
		\\&  \quad 0\leq\frac{1}{T} \sum_{t=0}^{T-1}  p_{k}(t) 
		\leq  \bar{P}_k.  \label{eq31c}
	\end{align}
\end{subequations}
Note that problem (\ref{eq31}) is a convex problem and satisfies the Slater's condition.
Thus, we can leverage the KKT conditions to obtain the optimal solution given in the following theorem.
\vspace{-0mm}
\begin{theorem}
	The optimal solution to problem \eqref{eq31}  is given as follows:
	\begin{itemize}
		\item
		If condition
		\vspace{-0mm}
		\begin{align}\label{eq32}
			\sum_{t=0}^{T-1}  \min \bigg\{\frac{\eta(t)}{|h_{k}(t)|^2},P_k^{\text{max}}\bigg\}
			\leq T \bar{P}_k
		\end{align}
		holds, then the optimal transmit power $p_{k}^{*}(t)$ is given by
		\begin{align}\label{eq33}
			p_{k}^{*}(t)
			=
			\min \bigg\{\frac{\eta(t)}{|h_{k}(t)|^2},P_k^{\text{max}}\bigg\}.
		\end{align}
		In this case, the transmit power either uses up the maximum power budget or has a form of channel inversion.
		\item Otherwise, the optimal transmit power $p_{k}^{*}(t)$ is given by
		\begin{align}\label{eq34}
			p_{k}^{*}(t) = \min \bigg\{\bigg(\frac{\sqrt{\eta(t)}|h_{k}(t)|}{|h_{k}(t)|^2+\mu_k^{*}\eta(t)}\bigg)^2,P_k^{\text{max}}\bigg\},
		\end{align}
		where $\mu^{*}_k$ can be found via the one-dimensional bisection search method to ensure that the average transmit power constraint $\sum_{t=0}^{T-1}  p_{k}^{*}(t)
		=T \bar{P}_k$ holds.
	\end{itemize}
	
\end{theorem}
\proof Please refer to Appendix \ref{proof3}. \qedhere

By now, problem $\mathscr{P}$ can be tackled by solving problems $\mathscr{P}_1$ and $\mathscr{P}_2$ alternately.
The proposed algorithm is summarized in {Algorithm \ref{alg1}}.

\begin{algorithm}[!t]
	\caption{Proposed alternating optimization algorithm for problem $\mathscr{P}$}
	\begin{algorithmic}[1]
		\STATE {\textbf{Input}}:   $\{h_{k}(t)\}_{t=0}^{T-1}$, stopping condition $\epsilon_0$.
		\STATE {{Initialize}}:  Transmit power $\{p_{k}(t)\}^0$ and  $i=0$.
		\REPEAT
		\STATE $i = i+1$.
		\STATE	Given  $\{p_{k}(t)\}^{i-1}$, update   $\{\eta(t)\}^{i}$  via
		$\{\eta(t)\}^{i} = \bigg(\frac{\sigma^2+\sum_{k=1}^K(\sqrt{\{p_{k}(t)\}^{i-1}}|h_{k}(t)|)^2}{\sum_{k=1}^K\sqrt{\{p_{k}(t)\}^{i-1}}|h_{k}(t)|}\bigg)^2.$
		\STATE Given $\{\eta(t)\}^{i}$,  update $\{p_{k}(t)\}^i$  via
		$	\{p_{k}(t)\}^i
		=
		\min \bigg\{\frac{\{\eta(t)\}^i}{|h_{k}(t)|^2},P_k^{\text{max}}\bigg\}$, if \eqref{eq32} holds
		or
		$\{p_{k}(t)\}^i = \min \bigg\{\bigg(\frac{\sqrt{\{\eta(t)\}^i}|h_{k}(t)|}{|h_{k}(t)|^2+\mu_k^{*}\{\eta(t)\}^i}\bigg)^2,P_k^{\text{max}}\bigg\}$, otherwise, where $\mu_k^{*}$ can be obtained via bisection search.
		\UNTIL $\frac{\overline{\sf {MSE}}^{i-1}-\overline{\sf {MSE}}^i}{\overline{\sf {MSE}}^i} < \epsilon_0$.
		\STATE {\textbf{Output}}:  $\{\eta(t)\}$ and $\{p_{k}(t)\}$. 
	\end{algorithmic}
	\label{alg1}	
\end{algorithm}

\begin{remark}
Although Algorithm \ref{alg1} can optimally solve problem $\mathscr{P}$, 
it has the following two limitations.
First, the computation complexity of Algorithm 1 is relatively high, 
as the iterative algorithm requires a large number of iterations to compute the transmit power of edge devices and the receive normalizing factor. 
Besides, one-dimensional bisection search is required in each iteration and hence introduces additional computation complexity.
Second, the alternating optimization algorithm requires the CSI of all communication rounds to solve problem $\mathscr{P}$.
However, it may not be practical to know the CSI of all communication rounds in advance, 
especially in time-varying wireless networks. 
To address these limitations, we shall propose an efficient knowledge-guided learning algorithm to solve problem $\mathscr{P}$ in the following section. 
\end{remark}

\section{Knowledge-Guided Learning Algorithm}
In this section, we propose an unsupervised learning algorithm and construct a DNN with domain knowledge to map the instantaneous CSI to the transmit power of edge devices and the receive normalizing factor. 
Instead of directly learning the mapping function,
the proposed learning algorithm leverages the structure of the optimal transmit power derived in Theorem 2 to enable low-complexity transceiver design for AirComp-assisted FL.

\subsection{Knowledge-Guided Learning for AirComp Transceiver Design}

We develop a knowledge-guided learning algorithm, which imitates the proposed alternating optimization algorithm in Section \ref{sec_opt}. 
In particular, the proposed learning algorithm learns a mapping between the instantaneous CSI of the current communication round and  the transmit power of edge devices and the receive normalizing factor by exploiting the structure information based on \eqref{eq34}.
The proposed neural network consists of multiple fully-connected layers and a structure mapping layer, as shown in Fig. 2. 
The fully-connected layers are designed to predict the dual variables (i.e., $\{\mu_k\}$) and the receive normalizing factor (i.e., $\eta$), 
while the  structure mapping layer after the fully-connected layers transforms the dual variables to the transmit powers of edge devices by exploiting the structure of the optimal transmit power derived in Section IV. 
 Specifically, in the structure mapping layer, the transmit powers $\{p_{k}\}$ of edge devices are generated according to the following structure 
\begin{align}\label{eq35}
	p_{k}=
	\min\biggl\{
	\bigg(\frac{\sqrt{\eta}|h_{k}|}{|h_{k}|^2+\mu_k\eta}\bigg)^2,P_k^{\text{max}}
	\biggr\}, \quad \forall \, k\in\mathcal{K}.
\end{align}

\begin{figure*}[!t]
	\centering
	\includegraphics[width=0.8\linewidth]{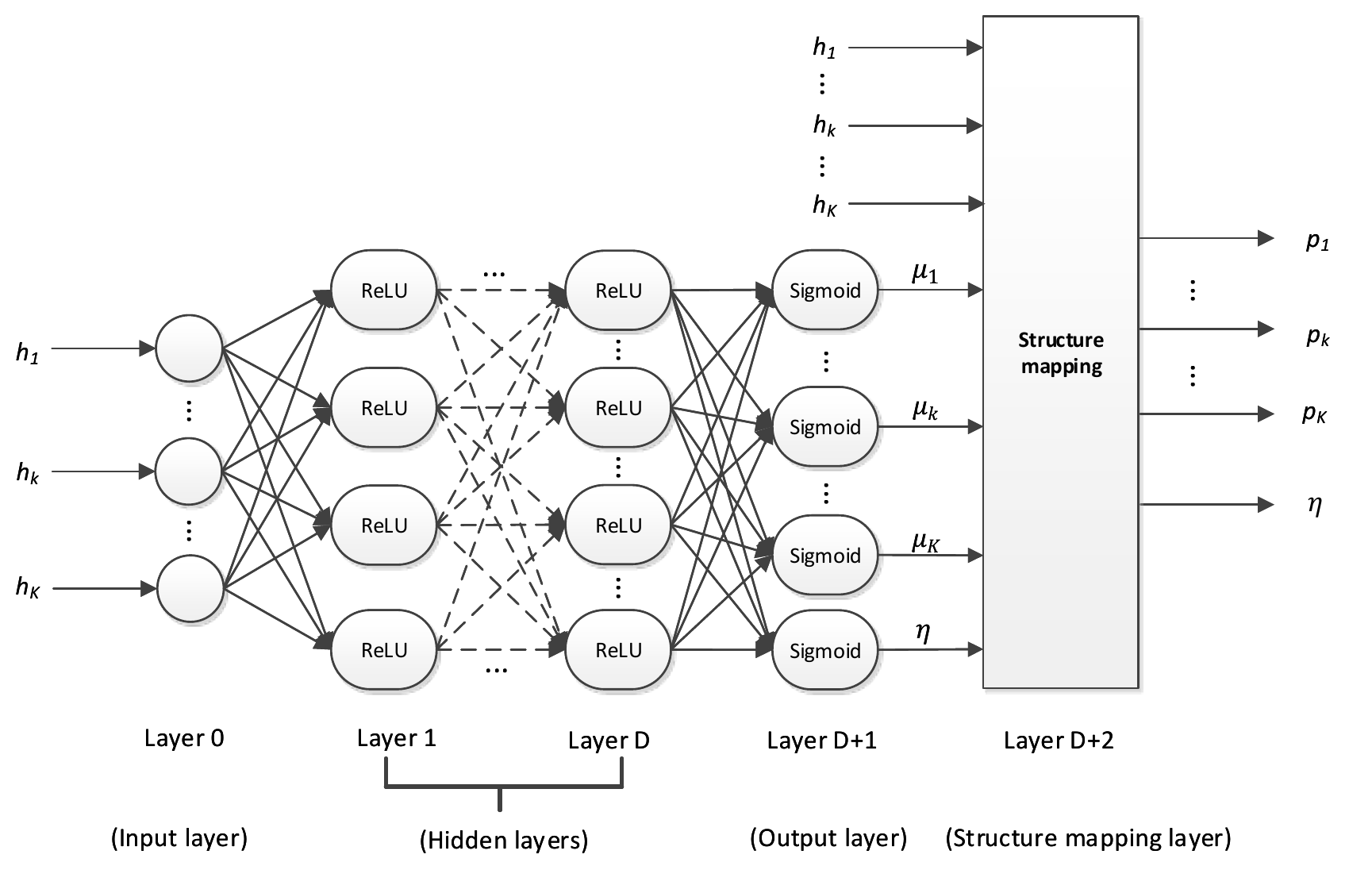}
	\vspace{-0.3cm}
	\caption{Architecture of the proposed knowledge-guided learning algorithm for the transceiver design of AirComp-assisted FL.}
	\label{DNN}
	\vspace{-0.5cm}
\end{figure*}

The structure mapping layer, which converts the dual variables and the receive normalizing factor to transmit powers of edge devices, is an important component of our proposed learning algorithm.
In particular, the  proposed learning algorithm, rather than directing estimating the transmit powers of edge devices, first predicts the dual variables that are key features extracted from the transmit power, 
and then utilizes the structure mapping layer to recover transmit power of  edge devices from the predicted dual variables.

The usage of the structure of the optimal transmit power enables the proposed neural network to efficiently find the optimal transmit power of edge devices and the optimal receive normalizing factor to minimize the signal misalignment error and the noise-induced error.
Compared  to the traditional fully-connected neural network that directly estimates the transmit power of edge devices, a salient feature of the proposed neural network is that the structure information is explicitly embedded into network architecture.
It is generally difficult for the conventional  fully-connected neural network to learn this structure because of the huge searching space of the transmit power.
By adopting the structure of the optimal transmit power, the searching space of the optimal transmit power can be reduced to a finite-dimensional subspace.  
In addition, the optimal transmit power is represented by dual variables $\mu_k$ and receive normalizing factor $\eta$.
Note that $\mu_k$ and $\eta$ are constrained, i.e., $\mu_k\geq 0$ and $\eta\geq 0 $. 
Hence, the finite-dimensional subspace can be further reduced, which in turn reduces the computation  complexity of searching for the optimal transmit power.
Although  the CSI of multiple communication rounds  is required to generate the training samples and to train the neural network, 
once the training process is completed, the proposed neural network only needs the CSI of the current  communication round to compute the transmit powers of edge devices and the receive normalizing factor.

\subsection{Deep Neural Network Design}
As shown in Fig. 2, the proposed neural network consists of an input layer, $D$ hidden layers, an output layer, and a structure mapping layer, which are indexed from $0$ to $D+2$.
\subsubsection{Input Layer}
The input layer has $K$ nodes corresponding to the channel coefficients between $K$ edge devices and the edge server.
The input layer, denoted as $\bm{z}_0$, can be expressed as $\bm{z}_0 = [h_{1},\ldots,h_{K}]^{\text{T}}$. 
\subsubsection{Hidden Layer} 
The hidden layers between the input layer and the output layer are fully connected.
We denote the output of the $d$-th neural layer as  $\bm{z}_d$ and  the number of nodes in the $d$-th layer  as $c_d$. 
We leverage the activation functions to generate  the estimated dual variables that are non-negative and continuous.
The output of the $d$-th layer is given by
\begin{align}
	\bm{z}_d = \text{ReLU}(\text{BN}(\bm{Q}_d\bm{z}_{d-1}+\bm{b}_d)), \quad d = 1,\ldots,D,
\end{align}
where $\bm{Q}_d$ is a weight matrix of dimension  $c_d\times c_{d-1}$, $\bm{b}_d$ is a vector of dimension $c_d$, $\text{ReLU}(x)$ denotes the ReLU function (i.e., $\max(x,0)$) that introduces nonlinearity, and $\text{BN}(\cdot)$ denotes the batch normalization layer.
The batch normalization layer is adopted to mitigate the sensitivity to the weight  initialization and to reduce the
probability of overfitting.

\subsubsection{Output Layer}
The dimension  of the output layer $c_{D+1}$ is set to $K+1$ and the output of the $(D+1)$-th layer is 
\begin{align}
	\bm{z}_{D+1} = \text{Sigmoid}(\bm{Q}_{D+1}\bm{z}_D+\bm{b}_{D+1}), 
\end{align}
where $\text{Sigmoid}(x) = \exp(x)/(1+\exp(x))$ denotes the sigmoid function. 

An intuitive approach is to directly train a fully-connected neural network that predicts the transmit power of edge devices and the receive normalizing  factor without a structure mapping layer.
However, such an  intuitive method cannot exploit the  specific structure of the optimal solution.
It is difficult for a fully-connected network to find  the optimal transmit power without using the structure information.
In contrast, we design the output of the output layer as the dual variables and the receive normalizing factor, i.e., 
\begin{align}
	\bm{z}_{D+1} = [\mu_1,\ldots,\mu_K,\eta]^{\text{T}}, 
\end{align}
where the first $K$ entries correspond to $K$ dual variables, and the last entry is the receive normalizing factor.
Vector $\bm{z}_{D+1}$ contains the key features of the transmit power of edge devices, and is passed through the structure mapping layer.

\subsubsection{Structure Mapping Layer}
In the structure mapping layer, the transmit powers $\{p_{k}\}$ of edge devices are calculated according to  \eqref{eq35}.
Adopting the structure information not only reduces the computation complexity to find the optimal transmit power, but also provides performance guarantee for the proposed learning algorithm.   

\vspace{-3mm}
\subsection{Deep Neural Network Training}
The conventional DL models are usually trained using supervised learning,  where the solution to the optimization problem is utilized as the ground truth and the squared error between the output of the neural network and the ground truth is the objective to be minimized. 
An obvious  drawback of the supervised learning based algorithm  is that a large number of labeled samples are required.
However, collecting the optimal solutions to  the optimization problem as labels is time-consuming. 
We in this paper adopt unsupervised learning, where only the CSI rather than the solutions of the optimization problem $\mathscr{P}$ is required as training samples.
The parameters of the neural network are optimized by using SGD.
Besides the neural network design, the loss function design is also important in our proposed algorithm.
We design our loss function as time-average MSE plus regularizer as follows
\begin{align}\label{loss}
	\text{loss} 
	= 
	\frac{1}{B}\sum_{t=1}^{B} \text{MSE}(t)
 	+ \gamma \cdot \text{Regularizer}, 
\end{align}
where $B$ is the size of training batch and $\gamma$ is the penalty parameter. 
To meet the average transmit power constraint, the regularizer that penalizes the constraint violation is defined as
\begin{align}
	&\text{Regularizer} = \sum_{k=1}^K \text{ReLU}\bigg(\frac{1}{B}\sum_{t=1}^B p_{k}(t)-\bar{P}_k\bigg).
\end{align} 
We adopt the ReLU function to ensure that the average transmit power is smaller than  $\bar{P}_k$ rigorously.
Note that the regularizer is not designed as  
$\sum_{k=1}^K \|\frac{1}{B}\sum_{t=1}^B p_{k}(t)-\bar{P}_k\|_2^2$
because $\ell_2$-norm can only ensure that the average transmit power to be close to $\bar{P}_k$.

\section{Simulation Results}
In this section, we present the simulation results of the proposed alternating optimization algorithm and knowledge-guided learning algorithm for AirComp-assisted FL.

\vspace{-3mm}
\subsection{Simulation Setup}\label{setup}
In the simulations, the wireless channels between the edge devices and the edge server over different communication rounds follow i.i.d. Rayleigh fading. 
The number of edge devices is set to  $K=20$ if not specified otherwise.
For each edge device, we define the receive signal-to-noise ratio (SNR) of device $k$ as  $\text{SNR}_k=\bar{P}_k/ \sigma^2$, which is set to $10\  \text{dB}$.
Besides, the maximum transmit power  is set to $P_k^{\text{max}}=3\bar{P}_k$. 
We describe the setting of other parameters  as follows.

\begin{itemize}
	\item[1)] \emph{Datasets:}
	We adopt non-i.i.d.  MNIST and CIFAR-10 datasets for FL training.
	In particular, we sort the data according to the labels, and divide the dataset into 200 shards with equal size.
	Each edge device is randomly assigned 2 data shards.
	\item[2)] \emph{FL neural network:}
	For the MNIST dataset, we train a convolutional neural network (CNN) with two convolution layers and two fully-connected layers.
	For the CIFAR-10 dataset, we train a CNN with three convolution layers and two fully-connected layers. 
	\item[3)] \emph{Knowledge-based neural network for AirComp transceiver design:}
	The number of the hidden layers (i.e., $D$) is $2$,
	while the numbers of nodes in two hidden layers are 256 and 64.
	\end{itemize}

	We compare the proposed alternating optimization algorithm and  knowledge-guided learning algorithm with the following four baseline methods:
	\begin{itemize}
	\item \textbf{Error-free transmission}: The  accumulated local gradients are assumed to be transmitted in an error-free manner, i.e., 
	without suffering from channel fading and receiver noise.
	The server receives each of the  accumulated local gradients from all edge devices without any distortion. 
	This benchmark characterizes the best FL performance.
	\item \textbf{Full power}: Each edge device transmits with a fixed power that is equal to the average transmit power budget $\bar{P}_k$.
	Besides, the receive normalizing factor is set  to 
	$\eta = \bigg(\frac{\sigma^2+\sum_{k=1}^K\bar{P}_k|h_k|^2}{\sum_{k=1}^K\sqrt{\bar{P}_k}|h_k|}\bigg)^2$.  
	\item \textbf{Channel inversion}: 
	Each edge device transmits with power defined as follows
	\begin{align}
		p_k = \left\{
		\begin{aligned}
			&	\min\left(\bar{P}_k,\frac{\eta}{|h_k|^2}\right),  && \bar{P}_k|h_k|^2 \geq \epsilon_{c}, \\
			& 	0 ,  && \bar{P}_k|h_k|^2 < \epsilon_c,
		\end{aligned}
		\right.
	\end{align}
	where $\eta=\min_k\bigg\{\frac{\sigma^2+ \bar{P}_k|h_k|^2}{ \sqrt{\bar{P}_k}|h_k|}\bigg\}$ and $\epsilon_c = 0.1$.
	\item \textbf{Knowledge-free learning}: A fully-connected neural network without structure information is trained to directly predict the transmit power of edge devices and the receive normalizing factor.
	Except for the structure mapping layer, the neural network structure is same as that of our proposed neural network.
	The output layer generates the  transmit power of edge devices and the receive normalizing factor, i.e., $
		\bm{z}_{D+1} = [p_1,\ldots,p_K,\eta]^{\text{T}}.$
	Finally, transmit power ${p_k}$ is multiplied by $\bar{P}_k$.
\end{itemize}
\begin{figure*}[!t]
	\centering
	\subfigure[MNIST dataset]{
		\includegraphics[width=0.45\linewidth]{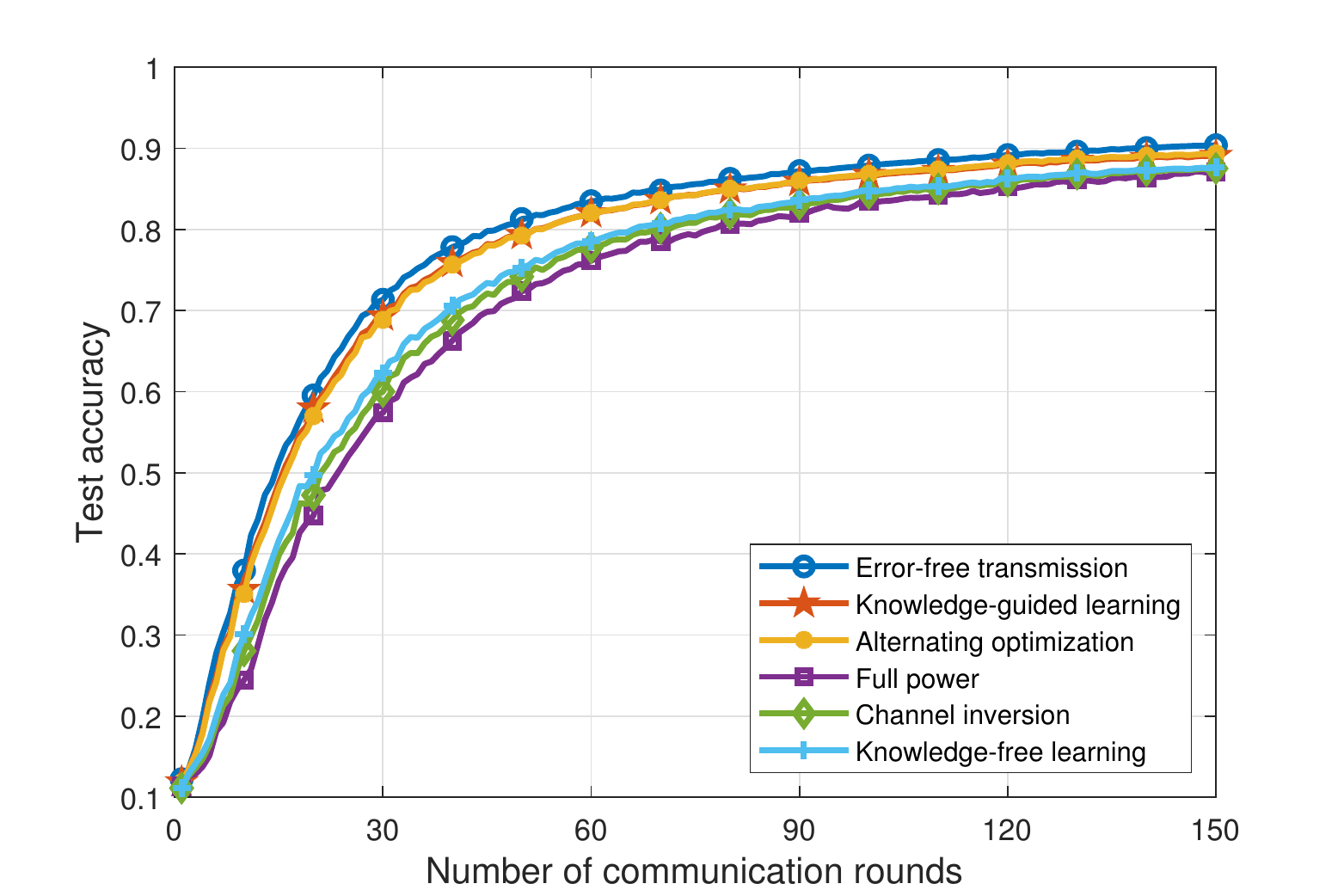}
	}
	\subfigure[MNIST dataset]{
		\includegraphics[width=0.45\linewidth]{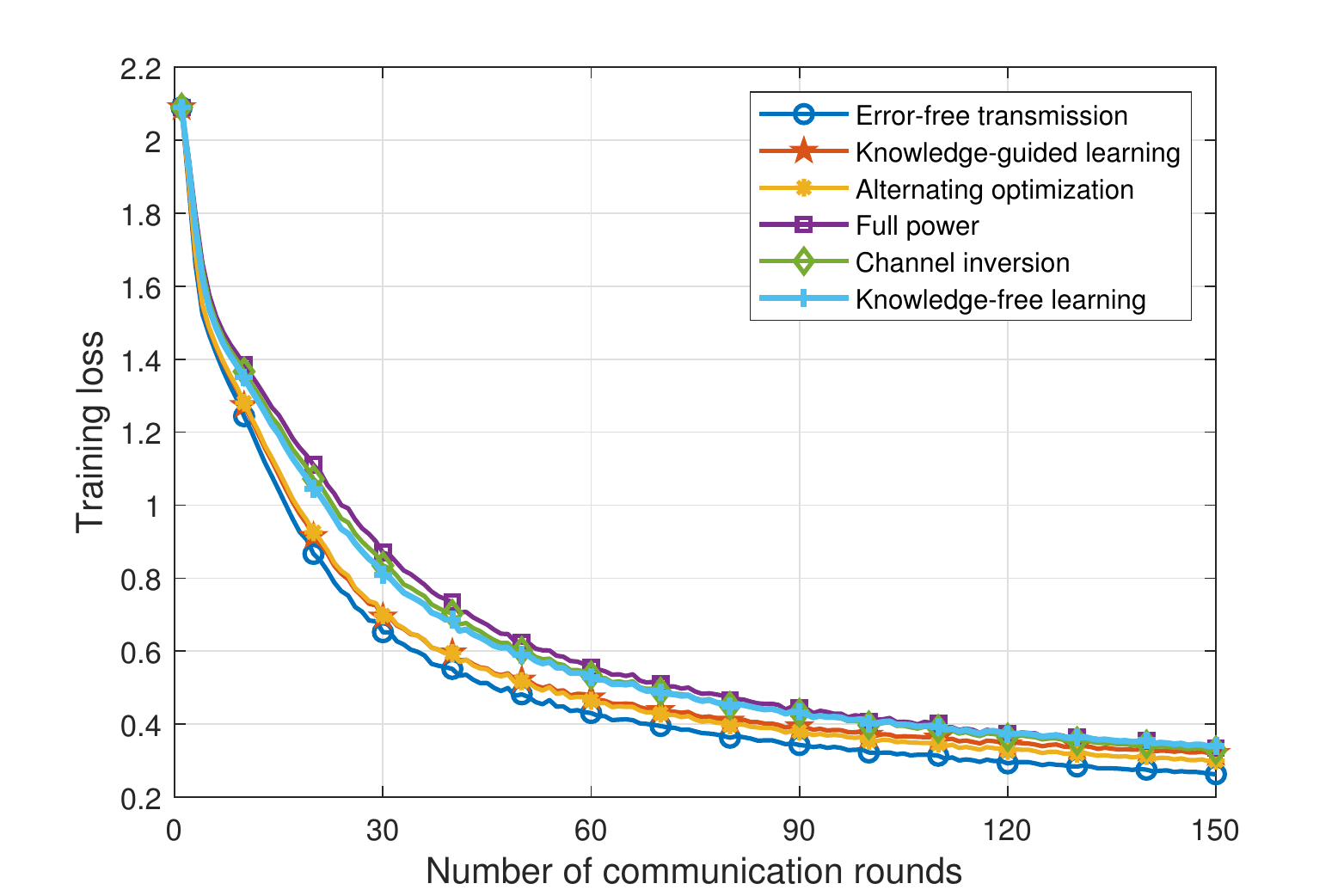}
	}
	\\
	\vspace{-0.38cm}
	\subfigure[CIFAR-10 dataset]{
		\includegraphics[width=0.45\linewidth]{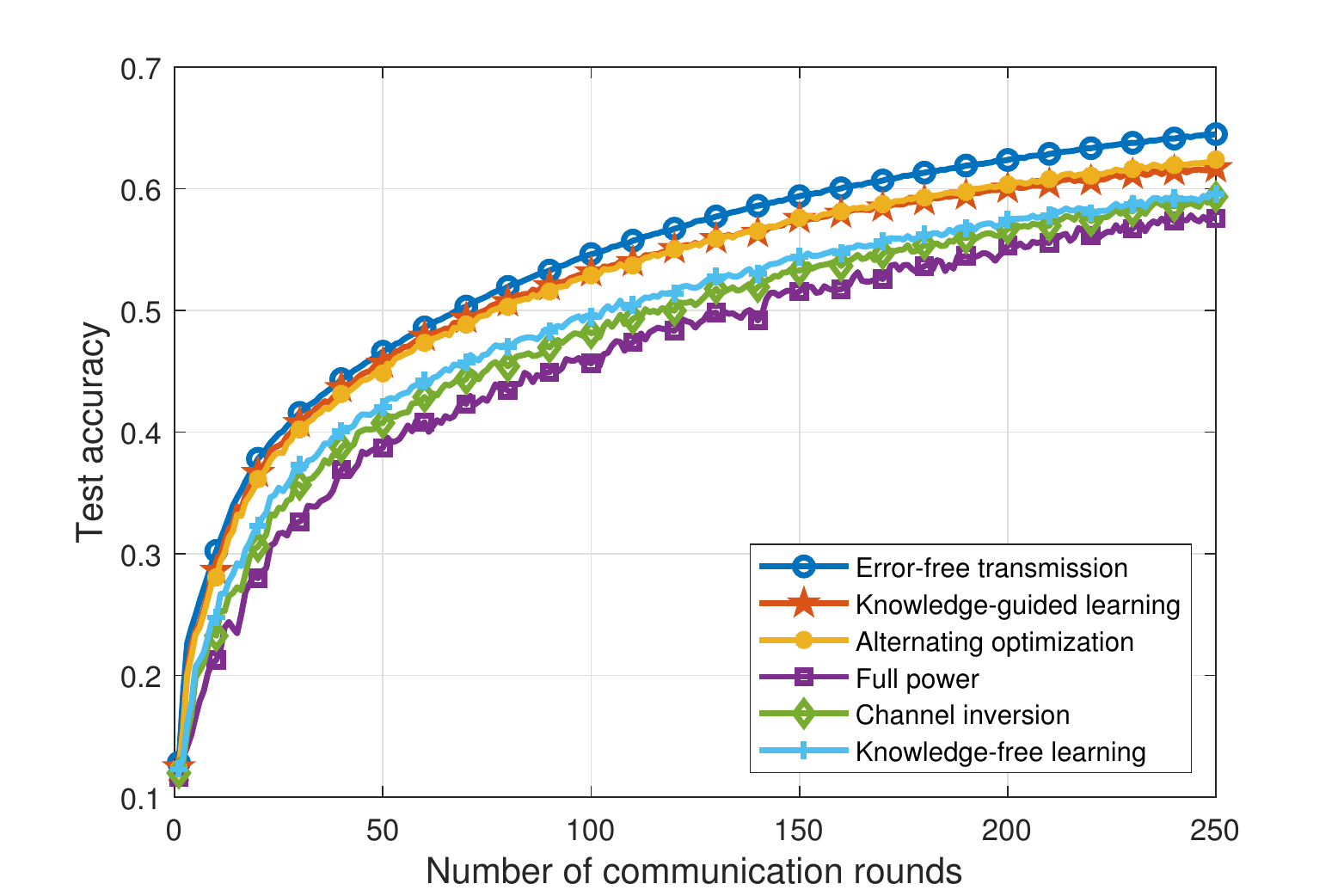}
	}
	\subfigure[CIFAR-10 dataset]{
		\includegraphics[width=0.45\linewidth]{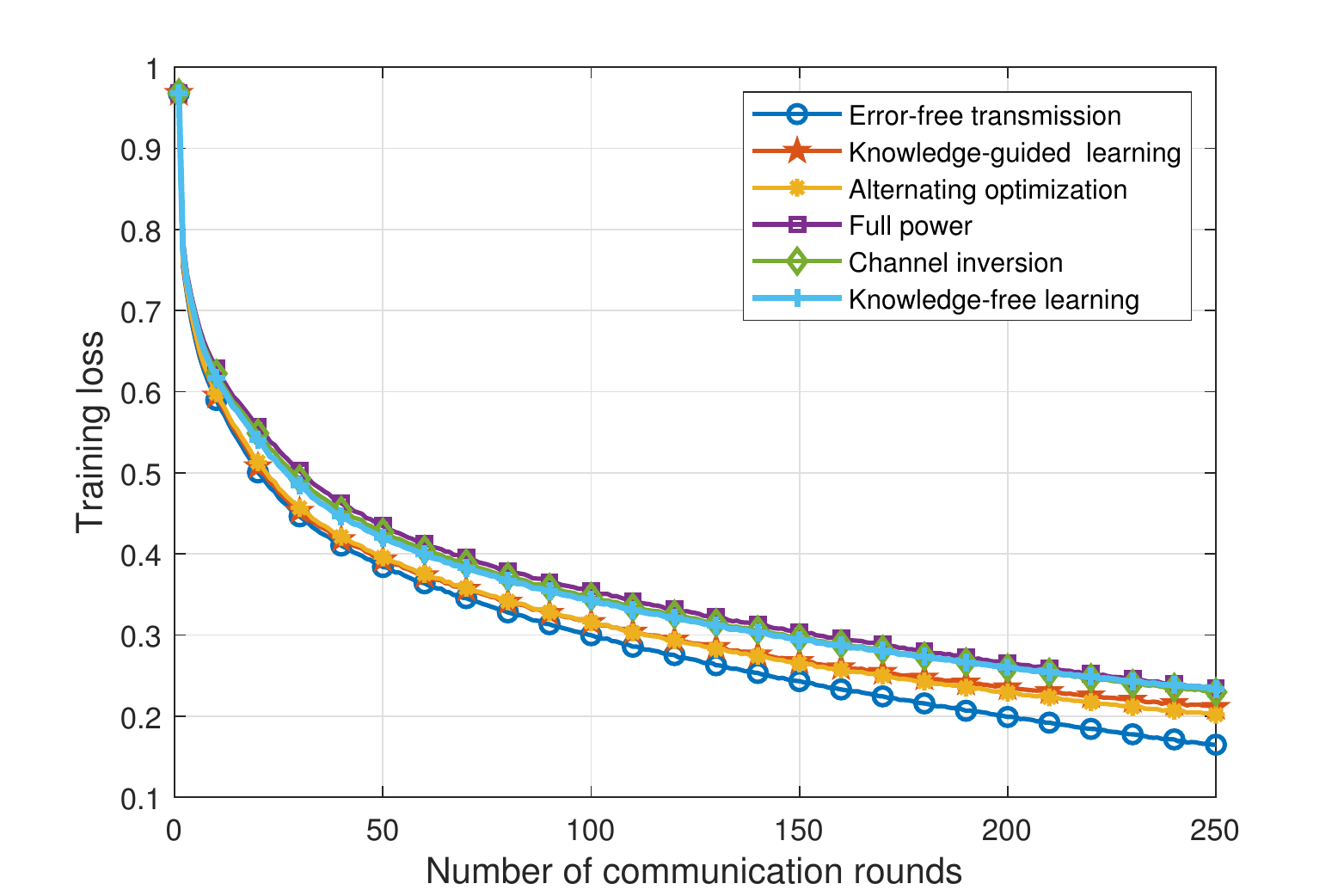}
	}
	\vspace{-0.1cm}
	\caption{Performance comparison of the proposed optimization and knowledge-guided algorithms with the baseline methods.}
	\vspace{-0.5cm}
\end{figure*}

\begin{figure*}[!t]
	\centering
	\subfigure[MNIST dataset ]{
		\includegraphics[width=0.45\linewidth]{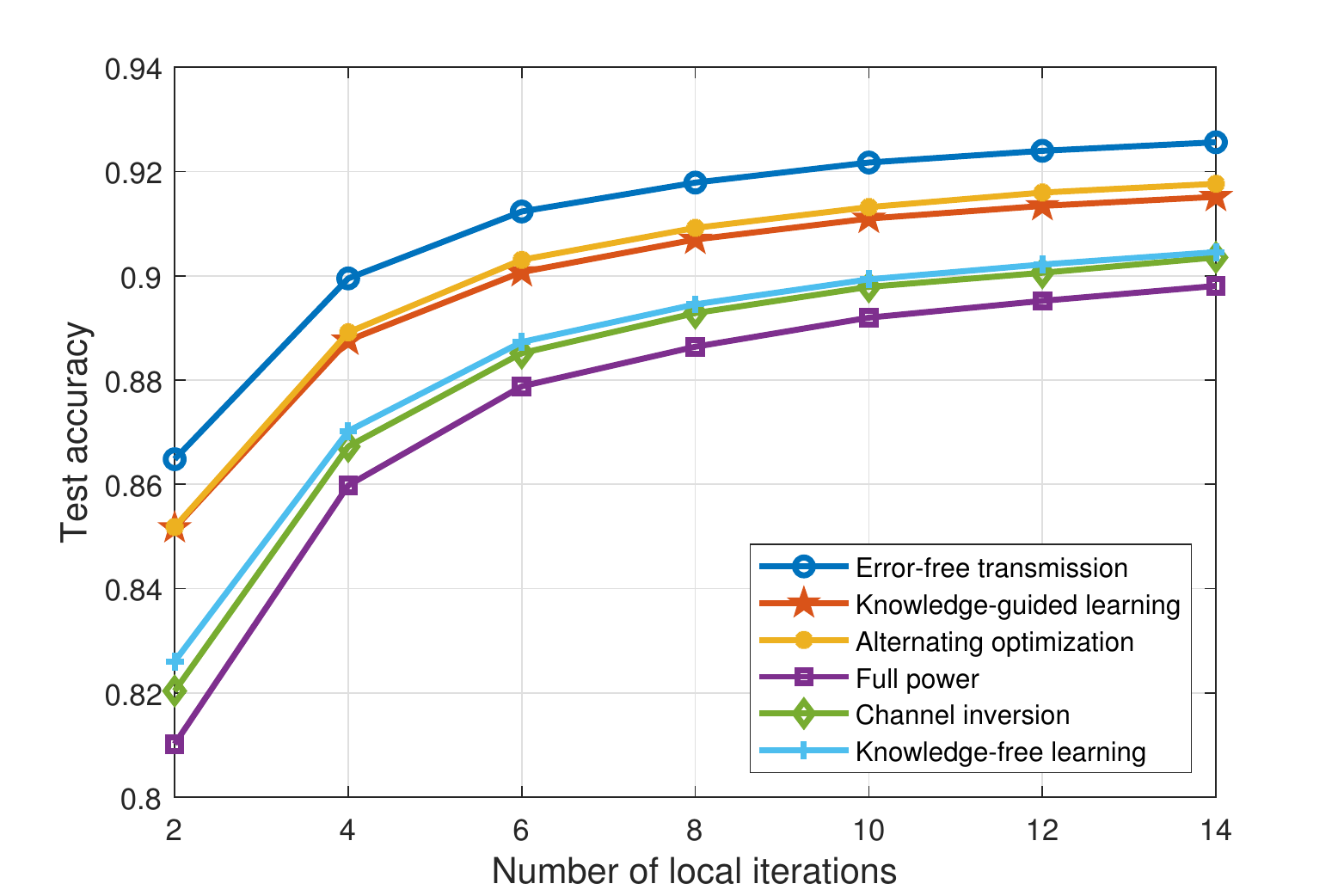}
	}
	\subfigure[MNIST dataset]{
		\includegraphics[width=0.45\linewidth]{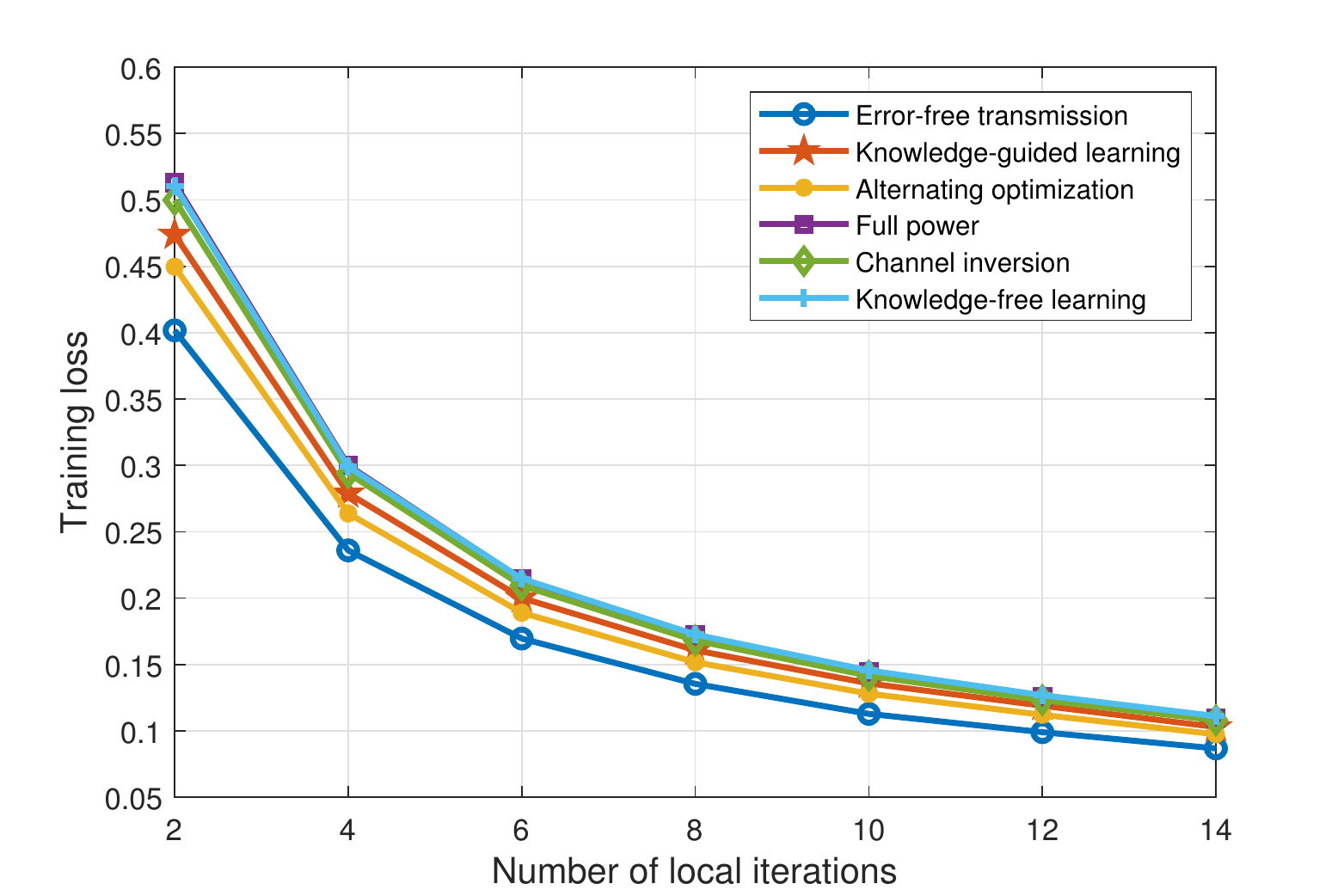}
	}
	\\
	\vspace{-0.38cm}
	\subfigure[CIFAR-10 dataset]{
		\includegraphics[width=0.45\linewidth]{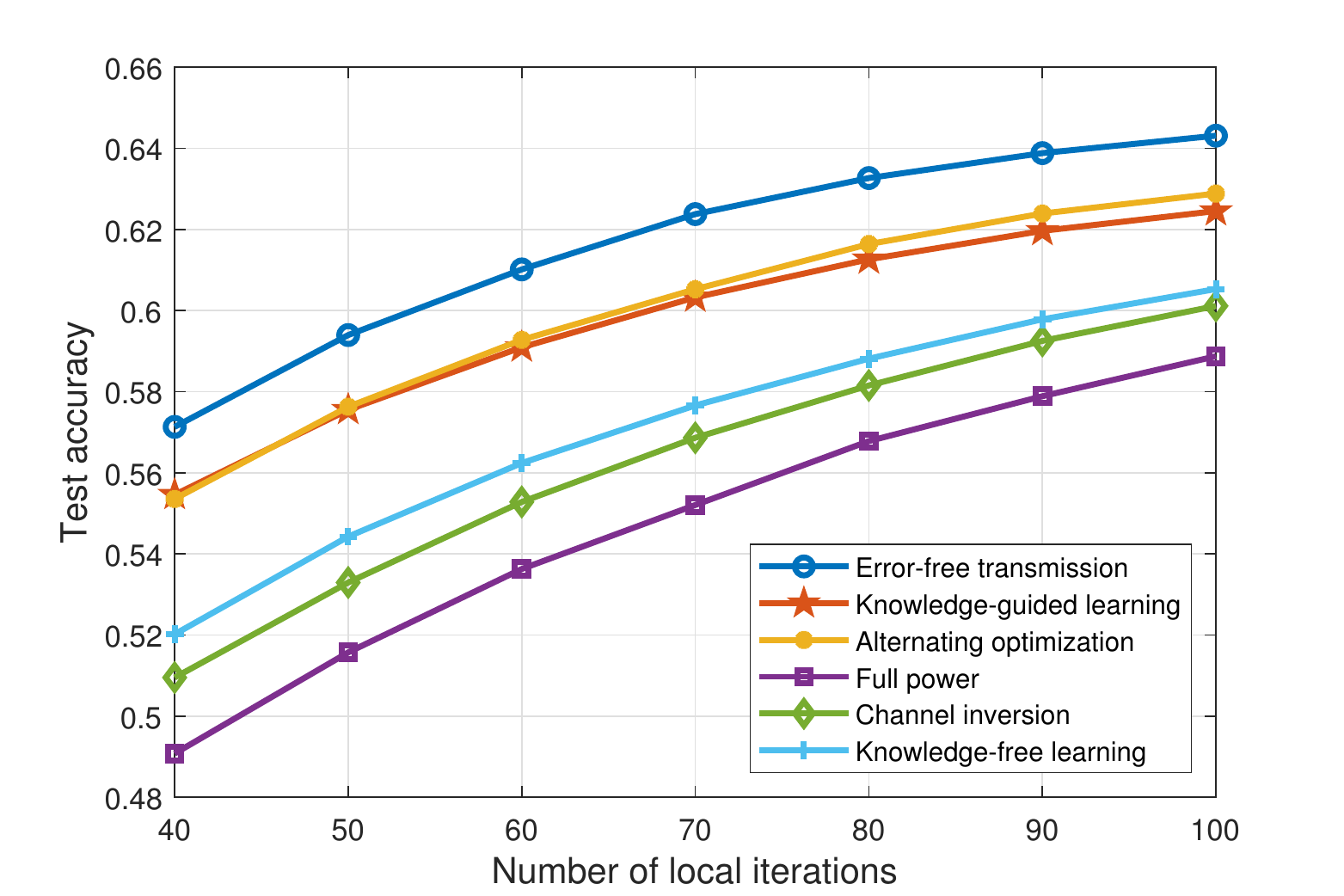}
	}
	\subfigure[CIFAR-10 dataset]{
		\includegraphics[width=0.45\linewidth]{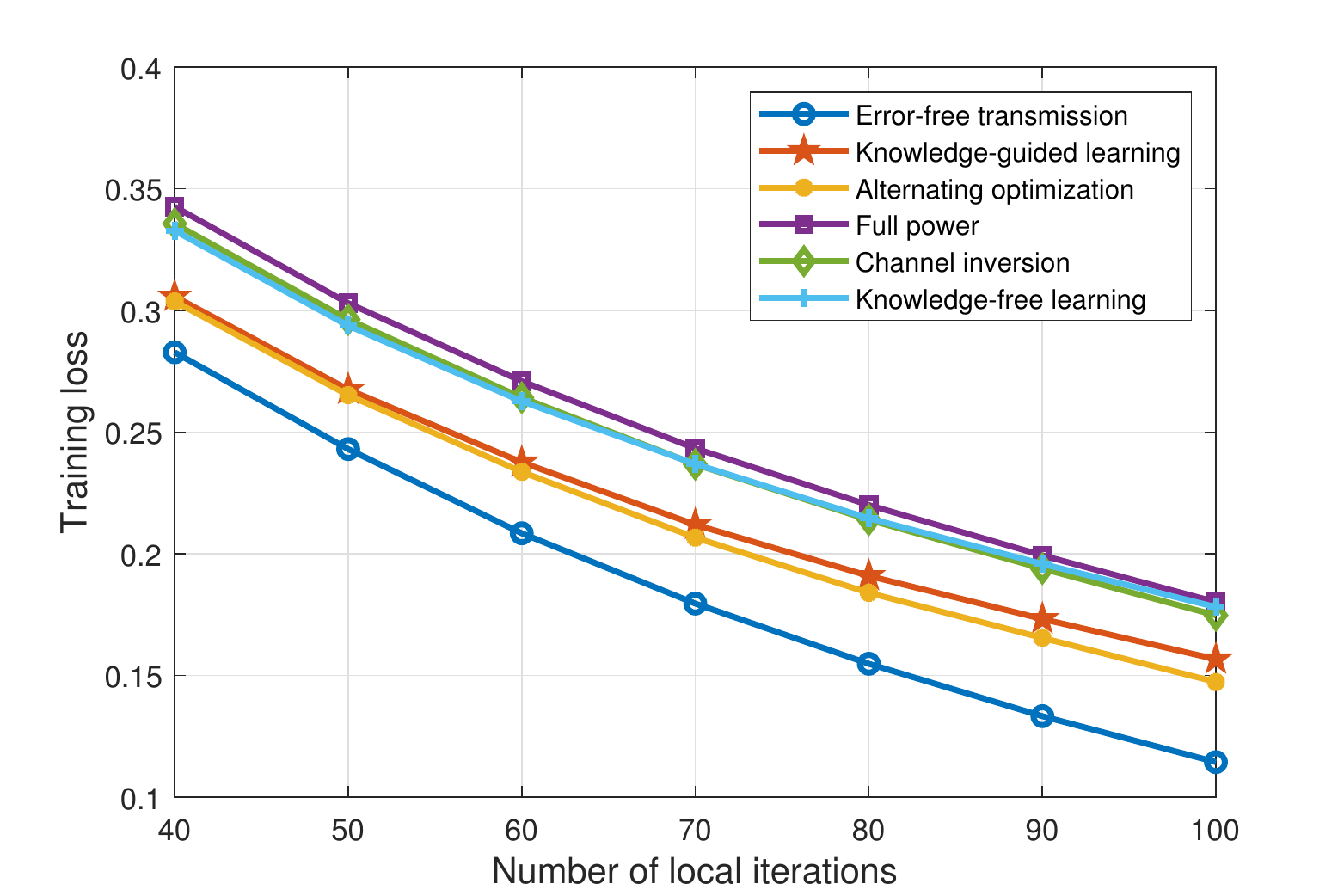}
	}
	\vspace{-0.1cm}
	\caption{Learning performance versus number of local iterations in terms of test accuracy and training loss.}
	\vspace{-0.5cm}
\end{figure*}

\vspace{-3mm}
\subsection{Performance Comparison}
Fig. 3 compares the test accuracy and the training loss of all schemes under consideration 
when the number of local iterations (i.e., $\phi$) is set to 3  and 50 for  MNIST and  CIFAR-10 datasets, respectively.
We observe that the error-free transmission achieves the best performance because of  the ideal model aggregation. 
The proposed  alternating optimization algorithm
obtains the optimal solutions of two subproblems and thus achieves the second-best convergence performance.
Besides, the performance gap between the alternating optimization algorithm and the knowledge-guided learning algorithm is quite small.
By exploiting  the structure information in terms of  the analytical expression of the optimal transmit power,
the proposed knowledge-guided learning algorithm can effectively optimize the transceiver design,
while  significantly reduces the  computation complexity.
Compared to the full power method, the channel inversion method, and the knowledge-free learning method, the proposed  knowledge-guided learning algorithm and the alternating optimization algorithm achieve faster convergence rates and better learning performance.
This demonstrates the importance of  the optimization of the transmit power and the receive normalizing factor as well as  the exploitation  of the structure information.
Similar performance trends in terms of training loss and test accuracy can be observed for all schemes under the MNIST and CIFAR-10 datasets.

Fig. 4 shows the learning performance of different algorithms with varying number of local iterations.
The  number of communication rounds (i.e., $T$) is set to 125 and 150 for MNIST and  CIFAR-10 datasets, respectively.
As shown in Fig. 4, the test accuracy increases with  the number of local iterations.
When the number of local iterations is large enough, the speed of performance increase becomes smaller. 
This is because excessive number of local iterations makes the local optimum deviate from the global mimimum
when the datasets  are non-i.i.d.
Besides, we observe that the knowledge-guided learning algorithm always achieves a performance close to that of the alternating optimization algorithm
on both MNIST and CIFAR-10 datasets.
By exploiting the structure of the optimal transmit power, 
the proposed knowledge-guided learning algorithm outperforms
the full power, channel inversion, and knowledge-free learning methods.

Fig. 5 shows the  learning performance of different algorithms  versus the  number of edge devices when
the number of communication rounds (i.e., $T$) and the number of local iterations (i.e., $\phi$) are set to $125$ and $2$, respectively.
We observe that the test accuracy of all schemes increases with the number of edge devices.
Specifically, the test accuracy increases rapidly when $K\leq20$, and  increases slowly when $K\geq25$.
This is because data redundancy occurs when too many edge devices are involved in FL training.
Fig. 5 also shows that the knowledge-guided learning algorithm is able to achieve comparable performance with the alternating optimzation algorithm under different number of edge devices.
In addition, the knowledge-guided learning algorithm and the alternating optimization algorithm significantly outperform the full power method, the channel-inversion method, and the knowledge-free learning method,
which clearly demonstrates the superiority of our proposed algorithms.

\begin{figure}[!t]
	\centering
	\includegraphics[width=0.5\linewidth]{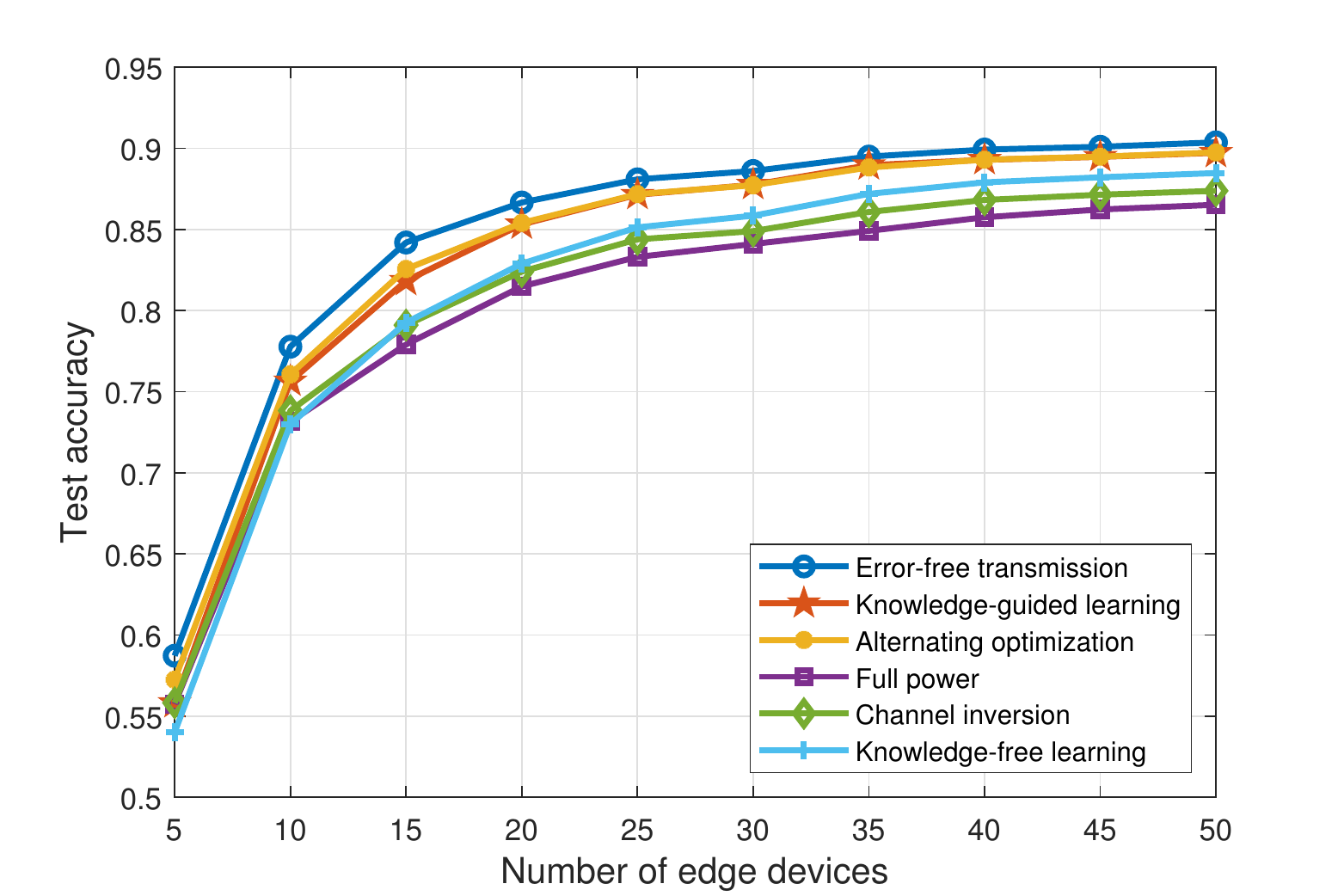}
	\vspace{-0.4cm}
	\caption{Test accuracy versus number  of edge devices for  MNIST dataset.}
	\label{systemmodel}
	\vspace{-1cm}
\end{figure}
\begin{minipage}[h]{\textwidth}
	\begin{minipage}[!t]{0.48\textwidth}
		\centering
		\makeatletter\def\@captype{table}\makeatother\caption{Computation time comparison and feasible probability of proposed knowledge-guided learning algorithm when  $T=200$}
		{
			\resizebox{\linewidth}{!}{
				\begin{tabular}{ccccccccc}
					\toprule  
					Number of edge devices &15&20&25&30&35\\
					\midrule  
					\tabincell{c}{Alternating optimization  } 
					& 44.63~s& 53.53~s &55.75~s & 58.71~s&59.40~s \\
					\tabincell{c}{Knowledge-guided learning  } 
					& 13.56~ms& 14.84~ms &17.15~ms & 20.95~ms&21.99~ms \\
					\tabincell{c}{Feasible probability of solutions } 
					& 100.00\%& 99.98\% &100.00\% & 99.99\%&100.00\% \\
					\bottomrule 
				\end{tabular}
			}
		}
	\end{minipage}
	\begin{minipage}[!t]{0.48\textwidth}
		\centering
		\makeatletter\def\@captype{table}\makeatother\caption{	Computation time comparison and feasible probability of proposed knowledge-guided learning algorithm when  $K=20$}
		{
			\resizebox{\linewidth}{!}{
				\begin{tabular}{cccccccccc}
					\toprule  
					\tabincell{c}{Number of  communication rounds} 
					&125&150&175&200&225&250\\
					\midrule  
					\tabincell{c}{Alternating optimization }
					& 31.22~s &39.98~s & 46.32~s&53.53~s &58.58~s &67.61~s \\
					\tabincell{c}{Knowledge-guided learning }  
					& 14.42~ms &15.37~ms & 14.83~ms&14.84~ms &15.67~ms &17.01~ms \\
					\tabincell{c}{Feasible probability  of solutions }
					& 99.71\% &99.66\% & 99.96\%&99.98\% &99.99\% &99.99\%\\
					\bottomrule 
				\end{tabular}
			}
		}
	\end{minipage}
\end{minipage}

\vspace{4mm}
We compare the computation time of the knowledge-guided learning algorithm and the alternating optimization algorithm, 
and evaluate the feasible  probability of the solutions returned by the knowledge-guided learning algorithm.
Table I shows the computation time 
versus the  number of edge devices
when the  number of communication rounds (i.e., $T$) is set to $200$.
We observe that  the computation time
grows as the number of edge devices increases.
The computation time of the knowledge-guided learning algorithm is three orders of magnitude smaller than the alternating optimization algorithm.
This is because the alternating optimization algorithm relies on an iterative process and each iteration involves  a bisection search, which are time-consuming.
Results demonstrate that the knowledge-guided learning algorithm is more  practical for the transceiver design of  AirComp-assisted FL with only marginal decrease in learning performance.
 
Table II shows that the computation time  when
the number of edge devices (i.e., $K$) is set to   $20$.
Compared with the alternating optimization algorithm, the knowledge-guided learning algorithm can speed up the computation by $2000-4000$ times.
Both Tables I and II show that the probability of feasible solution of the knowledge-guided learning algorithm is more than $99.6\%$ for  different number of communication rounds and different number of edge devices, which demonstrates the robustness of our proposed learning algorithm.

%

\section{Conclusion}

In this paper, we studied over-the-air FL,  taking in account multiple local SGD iterations and non-i.i.d. data. 
We derived the convergence of AirComp-assisted FL in terms of the time-average norm of gradients, followed by formulating an optimization problem to minimize the convergence bound. 
We first proposed an alternating optimization algorithm to obtain the optimal transmit power of edge devices and the receive normalizing factor, which, however, requires the global CSI and suffers from high computation complexity.
We further developed a knowledge-guided learning algorithm that exploits the domain knowledge to map the instantaneous CSI to the transmit power of edge devices and the receive normalizing factor.
Simulation results demonstrated
that the knowledge-guided learning algorithm  achieves a comparable performance as the alternating optimization algorithm, but with a much lower computation complexity.

\vspace{-4mm}
\appendix
\subsection{Proof of Theorem 1} \label{proof1}
Before proving Theorem 1, we first present the following four useful lemmas, which are proved  in Appendix C.
 
\vspace{-4mm}
\begin{lemma} \label{lemma1}
	With Assumption \ref{assumption3},   the difference between the global model vector and the individual local model vector is bounded, i.e., 
	\begin{align}
		&\mathbb{E} \bigg[ \frac{1}{K}\sum_{k=1}^K 
		\| \bm{w}(t)- \bm{w}_k(t,\zeta) \|_2^2\bigg] 
		\leq \phi\lambda^2  \xi^2 
		 +\phi \lambda^2 \sum_{q=0}^{\phi-1} 
		\chi \bigg\| \frac{1}{K}\sum_{k=1}^K \nabla F_k(\bm{w}_k(t,q)) \bigg\|_2^2.
	\end{align}
\end{lemma}
\vspace{-1mm}
\begin{lemma}\label{lemma2}
	With Assumptions \ref{assumption2} and \ref{assumption3}, the following equality holds  
	\begin{align}
		\mathbb{E}\left[ \left\langle \nabla F(\bm{w}(t)), \bm{\theta}(t)  \right\rangle\right]  
		 &\geq \frac{1}{2} \phi
		\| \nabla F(\bm{w}(t)) \|_2^2 
		 +
		\frac{1}{2}  \sum_{\zeta=0}^{\phi-1}  \bigg\|\frac{1}{K} \sum_{k=1}^K \nabla F_k(\bm{w}_k(t,\zeta)) \bigg\|_2^2  \notag
		\\&-  L^2  \frac{1}{2}  \sum_{\zeta=0}^{\phi-1} \frac{1}{K} \sum_{k=1}^K  \|( \bm{w}(t)-    \bm{w}_k(t,\zeta)) \|_2^2.
	\end{align}
\end{lemma}
\begin{lemma}\label{lemma3}
	With Assumption \ref{assumption3}, the average accumulated gradient norm is upper bounded as follows 
	\vspace{-1mm}
	\begin{align}\label{eq45}
		\mathbb{E} \left[\|\bm{\theta}(t)  \|_2^2\right]
		&\leq  \frac{\phi}{K}  \xi^2
		+ \phi \sum_{\zeta=0}^{\phi-1} \bigg\|  \frac{1}{K} \sum_{k=1}^K  \nabla F_k(\bm{w}_k(t,\zeta)) \bigg\|_2^2,
	\end{align}
	which relates the average accumulated gradient to the local full gradient. 
\end{lemma}
\vspace{-3mm}
\begin{lemma}\label{lemma4}
	With Assumptions \ref{assumption3} and \ref{assumption4}, the aggregation error and the instantaneous MSE has the following relationship  
	\vspace{-1mm}
	\begin{align}
		\mathbb{E}	[\| \bm{e}(t) \|_2^2]  
		\leq N \frac{\Gamma(K+1)}{K^2} \text{MSE}(t).
	\end{align}
\end{lemma}	
\vspace{-2mm}
\noindent\textbf{Proof of Theorem \ref{theorem}.}
According to Assumption \ref{assumption2}, $F(\bm{w})$ is $L$-smooth and we have the following inequality
\vspace{-2mm}
\begin{align} \label{eq47}
	& F(\bm{w}(t+1))-F(\bm{w}(t))  
	\leq -\lambda \left\langle \nabla F(\bm{w}(t)), \bm{\theta}(t) + \bm{e}(t) \right\rangle + \frac{\lambda^2 L}{2}  \|\bm{\theta}(t) + \bm{e}(t)  \|_2^2 \notag
	\\&= -\lambda \left\langle \nabla F(\bm{w}(t)), \bm{\theta}(t)   \right\rangle 
	-\lambda \left\langle \nabla F(\bm{w}(t)),  \bm{e}(t) \right\rangle 
	 +\frac{\lambda^2 L}{2} \|\bm{\theta}(t) \|_2^2 + \frac{\lambda^2 L}{2} \|\bm{e}(t)  \|_2^2
	+ \lambda^2 L \left\langle  \bm{\theta}(t),\bm{e}(t)  \right\rangle \notag
	\\&\mathop{\leq}^{(a)} \!-\!\lambda \left\langle\! \nabla F(\bm{w}(t)), \bm{\theta}(t) \! \right\rangle 
	\!+\! \frac{\lambda}{2} \|\!\nabla\! F(\bm{w}(t))\!\|_2^2 
	\!+\!  \frac{\lambda}{2} \|\bm{e}(t)\|_2^2  
	\!+\!\frac{\lambda^2 L}{2} \|\bm{\theta}(t)  \|_2^2 \!+\! \frac{\lambda^2 L}{2} \|\bm{e}(t)  \|_2^2
	\!+\! \lambda^2 L\! \left\langle \! \bm{\theta}(t),\bm{e}(t) \! \right\rangle \notag
	\\&\mathop{\leq}^{(b)} -\lambda \left\langle \nabla F(\bm{w}(t)), \bm{\theta}(t)  \right\rangle  
	+ \frac{\lambda}{2} \|\nabla F(\bm{w}(t))\|_2^2  
	 + (\frac{\lambda}{2}+\lambda^2 L) \|\bm{e}(t)\|_2^2 
	+\lambda^2 L \|\bm{\theta}(t) \|_2^2,
\end{align}
where $(a)$ follows from $-\bm{a}^T\bm{b}\leq \frac{\|\bm{a}\|_2^2}{2}+\frac{\|\bm{b}\|_2^2}{2}$ and $(b)$ follows by utilizing  $\bm{a}^T\bm{b}\leq \frac{\|\bm{a}\|_2^2}{2}+\frac{\|\bm{b}\|_2^2}{2}$.
By taking expectations over stochastic sampling and receiver noise at both sides of (\ref{eq47}), we obtain
\begin{align}
	\mathbb{E}[F(\bm{w}(t+1))-F(\bm{w}(t)) ] \notag
	&\leq -\lambda \mathbb{E} \left[\left[\left\langle \nabla F(\bm{w}(t)), \bm{\theta}(t)  \right\rangle \right]\right]
	\\&\!+\! \frac{\lambda}{2} \|\nabla F(\bm{w}(t))\|_2^2 
	 \!+ \! (\frac{\lambda}{2}+\lambda^2 L) \mathbb{E} [\|\bm{e}(t)\|_2^2]
	+\lambda^2 L \mathbb{E} [\|\bm{\theta}(t) \|_2^2].
\end{align}
Using Lemmas \ref{lemma1}, \ref{lemma2}, \ref{lemma3} and \ref{lemma4}, we have
\begin{align}\label{eq49}
&\mathbb{E}[F(\bm{w}(t+1))-F(\bm{w}(t)) ] 
= \frac{\lambda(1-\phi)}{2}
\| \nabla F(\bm{w}(t)) \|_2^2 
+ \bigg(\frac{\phi^2\lambda^3L^2}{2}+\frac{\phi\lambda^2L}{K}\bigg)\xi^2 
\notag
\\&+ (-1+\phi^2L^2\lambda^2\chi+2\phi\lambda L) \frac{\lambda}{2} \sum_{\zeta=0}^{\phi-1} \bigg\|  \frac{1}{K} \sum_{k=1}^K  \nabla
F_k(\bm{w}_k(t,\zeta)) \bigg\|_2^2 
+  (\frac{\lambda}{2}+\lambda^2 L) \mathbb{E} [\|\bm{e}(t)\|_2^2] \notag
\\&\mathop{\leq}^{(a)} \frac{\lambda(1-\phi)}{2}
\| \nabla F(\bm{w}(t)) \|_2^2 
+ \bigg(\frac{\phi^2\lambda^3L^2}{2}+\frac{\phi\lambda^2L}{K}\bigg)\xi^2 
+  (\frac{\lambda}{2}+\lambda^2 L) N \frac{\Gamma(K+1)}{K^2} \text{MSE}(t),
\end{align}
where (a) holds because $ \phi^2L^2\lambda^2\chi+2\phi\lambda L \leq 1$.
By summing up \eqref{eq49} for all $T$ communication rounds and rearranging the terms, we have
\begin{align}
	\mathbb{E}[F(\bm{w}(T))-F(\bm{w}(0))] 
	&\leq \frac{\lambda}{2}(1-\phi)\sum_{t=0}^{T-1}\|\nabla F(\bm{w}(t))\|_2^2 \notag
	\\&+ \!\bigg(\!\frac{\phi^2\lambda^3L^2}{2}\!+\!\frac{\phi\lambda^2L}{K}\!\bigg)\xi^2  
	\!+\!  (\frac{\lambda}{2}\!+\!\lambda^2 L)  \!\sum_{t=0}^{T-1}\! N \frac{\Gamma(K+1)}{K^2} \text{MSE}(t).
\end{align}
With Assumption \ref{assumption1}, we have 
$
	F(\bm{w}(T))-F(\bm{w}(0)) \geq F(\bm{w}^{*})-F(\bm{w}(0)),
$
which yields \eqref{eq21}.

\subsection{Proof of Theorem 2}\label{proof3}

The Lagrangian function of problem (\ref{eq31}) is given by
\begin{align}
&\mathcal{L}\!(\!\{p_{k}(t)\}\!,\!\{\alpha(t)\}\!,\!\mu_k\!) \!
\!=\! \sum_{t=0}^{T-1} \! \left(\!\frac{\sqrt{p_{k}(t)}|h_{k}(t)|}{\sqrt{\eta(t)}}\!-1\!\right)^2 \!
+\! \sum_{t=0}^{T-1}\!\alpha(t)\!\bigg(\!p_{k}(t)\!-\!P_k^{\text{max}}\!\bigg)\!
+\! \mu_k\bigg(\!\sum_{t=0}^{T-1}p_{k}(t)\!-\!T\bar{P}_k\!\bigg)\!,\notag
\end{align}
where $\{\alpha(t)\geq 0 \}$ denote the dual variables associated with the constraints in \eqref{eq31b} and $\mu_k\geq 0$ denotes the dual variable associated with constraint \eqref{eq31c}.
By setting the first derivative of $\mathcal{L}(\{p_{k}(t)\},\{\alpha(t)\},\mu_k)$ with respect to $p_{k}(t)$ to zero as follows
\begin{align}
	\frac{\partial\mathcal{L}}{\partial p_{k}(t)} = \frac{|h_{k}(t)|^2}{\eta(t)} - \frac{|h_{k}(t)|}{\sqrt{\eta(t)p_{k}(t)}} + \alpha(t) + \mu_k = 0,
\end{align}
we obtain
\begin{align}
	p_{k}(t)=\bigg(\frac{\sqrt{\eta(t)}|h_{k}(t)|}{|h_{k}(t)|^2+(\alpha(t)+\mu_k)\eta(t)}\bigg)^2.
\end{align}

We denote $\{p_{k}^{*}(t)\}$ as the optimal transmit power, and $\{\alpha^{*}(t)\}$ and $\mu_k^{*}$ as the optimal dual variables.
The optimal $\{p_{k}^{*}(t)\}$, $\{\alpha^{*}(t)\}$,  and $\mu_k^{*}$ should satisfy the following KKT conditions 
\begin{align}
	&p_{k}^{*}(t)=\bigg(\frac{\sqrt{\eta(t)}|h_{k}(t)|}{|h_{k}(t)|^2+(\alpha^{*}(t)+\mu_k^{*})\eta(t)}\bigg)^2,\forall\, t, \label{eq65}
	\\&0\leq  p_{k}^{*}(t)\leq P_k^{\text{max}},
	\forall\, t, \label{eq66}
	\\&0\leq \sum_{t=0}^{T-1}   p_{k}^{*}(t) 
	\leq T\bar{P}_k,  \label{eq67}
	\\&\alpha^{*}(t)\geq 0,\forall\, t, \label{eq68}
	\\&\mu_k^{*} \geq 0, \label{eq69}
	\\&\alpha^{*}(t)\bigg(p_{k}^{*}(t)-P_k^{\text{max}}\bigg)=0,  \forall\, t, \label{eq70}
	\\&\mu_k^{*}\bigg(\sum_{t=0}^{T-1}\  p_{k}^{*}(t)
	- T \bar{P}_k \bigg)=0. \label{eq71}
\end{align}

If $\alpha^{*}(t)>0$, because of  the comlementary slackness condition (\ref{eq70}), then we obtain
$
	p_{k}^{*}(t) = P_k^{\text{max}}.
$
If $\alpha^{*}(t)=0$, then we obtain
$
	p_{k}^{*}(t)=\bigg(\frac{\sqrt{\eta(t)}|h_{k}(t)|}{|h_{k}(t)|^2+\mu_k^{*}\eta(t)}\bigg)^2.
$
Hence, the analytical expression of $p_{k}^{*}(t)$ is given by
\begin{align}\label{eq74}
	p_{k}^{*}(t)=\left\{
	\begin{aligned}
		&
		\bigg(\frac{\sqrt{\eta(t)}|h_{k}(t)|}{|h_{k}(t)|^2+\mu_k^{*}\eta(t)}\bigg)^2, &\text{if}\  \alpha^{*}(t)=0	
		,
		\\&P_k^{\text{max}}, &\text{if}\  \alpha^{*}(t)>0.
	\end{aligned}
	\right.
\end{align}
As the value of $p_{k}^{*}(t)$ depends on $\alpha^{*}(t)$, we discuss two following cases. 
If $\bigg(\frac{\sqrt{\eta(t)}|h_{k}(t)|}{|h_{k}(t)|^2+\mu_k^{*}\eta(t)}\bigg)^2 > P_k^{\text{max}}$ and $\alpha^{*}(t)=0$, then
\begin{align}
	p_{k}^{*}(t)\mathop{=}^{(a)}\bigg(\frac{\sqrt{\eta(t)}|h_{k}(t)|}{|h_{k}(t)|^2+\mu_k^{*}\eta(t)}\bigg)^2 
	>
	P_k^{\text{max}},
\end{align}
where $(a)$ is due to \eqref{eq74}.
Note that the maximum power constraint (\ref{eq66}) is not satisfied.
Hence, $\alpha^{*}(t)>0$ must hold, resulting in $p_{k}^{*}(t) = P_k^{\text{max}}$.
If $\bigg(\frac{\sqrt{\eta(t)}|h_{k}(t)|}{|h_{k}(t)|^2+\mu_k^{*}\eta(t)}\bigg)^2 \leq P_k^{\text{max}}$ and $\alpha^{*}(t)>0$,
then 
\begin{align}
	p_k^{*}(t) &\mathop{=}^{(a)} P_k^{\text{max}}
	\geq \bigg(\frac{\sqrt{\eta(t)}|h_{k}(t)|}{|h_{k}(t)|^2+\mu_k^{*}\eta(t)}\bigg)^2 
	\mathop{>}^{(b)}
	\bigg(\frac{\sqrt{\eta(t)}|h_{k}(t)|}{|h_{k}(t)|^2+(\alpha^{*}(t)+\mu_k^{*})\eta(t)}\bigg)^2,
\end{align}
where $(a)$ is due to \eqref{eq74} and $(b)$ is due to $\alpha^{*}(t)>0$.
Note that \eqref{eq65} is not satisfied.
Hence, $\alpha^{*}(t)=0$ must holds, which leads to $p_k^{*}(t) = \bigg(\frac{\sqrt{\eta(t)}|h_{k}(t)|}{|h_{k}(t)|^2+\mu_k^{*}\eta(t)}\bigg)^2$. 
To sum up, the value of $p_{k}^{*}(t)$ is independent of  $\alpha^{*}(t)$ and is given by $p_{k}^{*}(t) = \min\bigg\{\bigg(\frac{\sqrt{\eta(t)}|h_{k}(t)|}{|h_{k}(t)|^2+\mu_k^{*}\eta(t)}\bigg)^2,P_k^{\text{max}}\bigg\}$.

If $\mu_k^{*}>0$, because of the complementary slackness condition (\ref{eq71}), then we obtain
$
	\sum_{t=0}^{T-1}  p_{k}^{*}(t)
	= T \bar{P}_k.
$
Hence, we obtain
\eqref{eq34},
where $\mu_k^{*}$ can be found to ensure $\sum_{t=0}^{T-1} p_{k}^{*}(t)
= T \bar{P}_k$ by using the one-dimensional bisection search method.
Furthermore, if $\mu_k^{*}=0$ and $\sum_{t=0}^{T-1}  p_{k,m}^{*}(t)
\leq T \bar{P}_k$, then we obtain
(\ref{eq33}).
If $\mu^{*}=0$ and $\sum_{t=0}^{T-1} p_{k}^{*}(t)
> T \bar{P}_k$, then (\ref{eq67}) does not hold.
Hence, $\mu_k^{*}>0$ must holds, which leads to  (\ref{eq34}).

To sum up, the optimal transmit power is given by
\begin{align}
p_{k}^{*}(t) =\left\{
\begin{aligned}
&\min \bigg\{\frac{\eta(t)}{|h_{k}(t)|^2},P_k^{\text{max}}\bigg\}, 
&  \text{if}\quad \sum_{t=0}^{T-1}  \min \bigg\{\frac{\eta(t)}{|h_{k}(t)|^2},P_k^{\text{max}}\bigg\}
\leq T \bar{P}_k, 
\\
&\min \bigg\{\bigg(\frac{\sqrt{\eta(t)}|h_{k}(t)|}{|h_{k}(t)|^2+\mu_k^{*}\eta(t)}\bigg)^2,P_k^{\text{max}}\bigg\},  
&  \text{if}\quad \sum_{t=0}^{T-1}  \min \bigg\{\frac{\eta(t)}{|h_{k}(t)|^2},P_k^{\text{max}}\bigg\}
>T \bar{P}_k.
\end{aligned}
\right.
\end{align}
where $\mu^{*}$ can be found to ensure the average power constraint $\sum_{t=0}^{T-1}  p_{k}^{*}(t)
=T \bar{P}_k$ via the one-dimensional bisection search method.

\subsection{Proof of Lemmas}
\noindent\textbf{Proof of Lemma 1.}
According to  \eqref{eq04}, the local model vector is given by 
$
	\bm{w}_k(t,\zeta) =  \bm{w}(t) - \lambda \sum_{q=0}^{\zeta-1} \tilde{\bm{g}}_k(t,q).
$
We bound $\mathbb{E} \bigg[ \frac{1}{K}\sum_{k=1}^K 
\| \bm{w}(t)- \bm{w}_k(t,\zeta) \|_2^2\bigg]$ as follows
\begin{align}
&\mathbb{E} \bigg[ \frac{1}{K}\sum_{k=1}^K 
\| \bm{w}(t)- \bm{w}_k(t,\zeta) \|_2^2\bigg] \notag
\!=\! \mathbb{E}\!\bigg[\! \frac{1}{K}\!\sum_{k=1}^K\!  \bigg\| \lambda \sum_{q=0}^{\zeta-1} \tilde{\bm{g}}_k(t,q) \bigg\|_2^2\bigg] \notag
\!=\! \frac{1}{K}\sum_{k=1}^K  \lambda^2  \mathbb{E}\bigg[\bigg\|  \sum_{q=0}^{\zeta-1} \tilde{\bm{g}}_k(t,q) \bigg\|_2^2\bigg] \notag
\\&\!\mathop{=}^{(a)}\! \frac{1}{K}\!\sum_{k=1}^K\!  \lambda^2
\text{Var}\!\bigg(\!\sum_{q=0}^{\zeta-1} \tilde{\bm{g}}_k(t,q)\!\bigg)\!
+ \!\frac{1}{K}\!\sum_{k=1}^K \! \lambda^2  \bigg\|\!\sum_{q=0}^{\zeta-1}\! \nabla F_k(\bm{w}_k(t,q)) \!\bigg\|_2^2 \notag
\!\mathop{=}^{(b)}\!  \frac{1}{K}\sum_{k=1}^K  \lambda^2 \!\sum_{q=0}^{\zeta-1}\! \text{Var}\bigg(\!\tilde{\bm{g}}_k(t,q)\!\bigg)
\\&\!+\! \frac{1}{K}\!\sum_{k=1}^K \! \lambda^2\! \bigg\|\!  \sum_{q=0}^{\zeta-1} \nabla F_k(\bm{w}_k(t,q))\! \bigg\|_2^2 \notag
\!\leq\! \frac{1}{K}\!\sum_{k=1}^K\!  \lambda^2\! \sum_{q=0}^{\zeta-1}\! \text{Var}\bigg(\!\tilde{\bm{g}}_k(t,q)\!\bigg)
\!+\! \frac{1}{K}\!\sum_{k=1}^K \! \lambda^2 \zeta \!\sum_{q=0}^{\zeta-1} \!\| \!  \nabla F_k(\bm{w}_k(t,q))\! \|_2^2 \notag
\\&\!\leq\! \frac{1}{K}\!\sum_{k=1}^K \! \lambda^2\! \sum_{q=0}^{\phi-1}\! \text{Var}\bigg(\!\tilde{\bm{g}}_k(t,q)\!\bigg) \!
+ \!\frac{1}{K}\!\sum_{k=1}^K\!  \phi\! \lambda^2 \!\sum_{q=0}^{\phi-1} \!\|\! \nabla F_k(\bm{w}_k(t,q)) \!\|_2^2 \notag
\\&\leq \lambda^2 \phi \xi^2 
+   \phi \lambda^2 \sum_{q=0}^{\phi-1} \frac{1}{K}\sum_{k=1}^K
\| \nabla F_k(\bm{w}_k(t,q)) \|_2^2
\mathop{\leq}^{(c)} \phi\lambda^2  \xi^2 
+   \phi \lambda^2 \sum_{q=0}^{\phi-1} 
\chi \bigg\| \frac{1}{K}\sum_{k=1}^K \nabla F_k(\bm{w}_k(t,q)) \bigg\|_2^2,
\end{align}
where  $(a)$ follows from  $\mathbb{E}[\bm{x}^2] = \text{Var}[\bm{x}] + [\mathbb{E}[\bm{x}]]^2$ and Assumption \ref{assumption3}, $(b)$ holds because  $\text{Var}(\sum_{j=1}^{n}\bm{x}_j)=\sum_{j=1}^n \text{Var}(\bm{x}_j)$ with independent $\{ \bm{x}_j \}$, 
and $(c)$ follows from  Definition \ref{definition1}.

\noindent\textbf{Proof of Lemma 2.}
Recall the definitions of $\bm{\theta}(t)$ and $\bm{\theta}_k(t)$, we have $\bm{\theta}(t) =\frac{1}{K}\sum_{k=1}^K \bm{\theta}_k(t)=  \frac{1}{K}\sum_{k=1}^K \sum_{\zeta=0}^{\phi-1} \tilde{\bm{g}}_k(t,\zeta) $.
Thus, we have
\begin{align}
&\mathbb{E}\bigg[ \bigg\langle \nabla F(\bm{w}(t)), \bm{\theta}(t)  \bigg\rangle\bigg]  \notag
\mathop{=}^{(a)}   \bigg\langle \nabla F(\bm{w}(t)),\frac{1}{K}\sum_{k=1}^K \sum_{\zeta=0}^{\phi-1} \nabla F_k(\bm{w}_k(t,\zeta))   \bigg\rangle \notag
\\&=   \sum_{\zeta=0}^{\phi-1}   \bigg\langle \nabla F(\bm{w}(t)), \frac{1}{K} \sum_{k=1}^K  \nabla F_k(\bm{w}_k(t,\zeta))   \bigg\rangle \notag
\mathop{=}^{(b)} 
\frac{1}{2}  \sum_{\zeta=0}^{\phi-1} 
\bigg[   \| \nabla F(\bm{w}(t)) \|_2^2 + \bigg\|\frac{1}{K} \sum_{k=1}^K \nabla F_k(\bm{w}_k(t,\zeta)) \bigg\|_2^2  \notag
\\&- \bigg\| \frac{1}{K} \sum_{k=1}^K (\nabla F_k(\bm{w}(t))-   \nabla F_k(\bm{w}_k(t,\zeta))) \bigg\|_2^2\bigg] \notag
\geq \frac{1}{2}  \sum_{\zeta=0}^{\phi-1} 
\bigg[   \| \nabla F(\bm{w}(t)) \|_2^2 + \bigg\|\frac{1}{K} \sum_{k=1}^K \nabla F_k(\bm{w}_k(t,\zeta)) \bigg\|_2^2  \notag
\\&-  \frac{1}{K} \sum_{k=1}^K  \|(\nabla F_k(\bm{w}(t))-   \nabla F_k(\bm{w}_k(t,\zeta))) \|_2^2\bigg] \notag
\mathop{\geq}^{(c)} \frac{1}{2}  \sum_{\zeta=0}^{\phi-1} 
\bigg[   \| \nabla F(\bm{w}(t)) \|_2^2 + \bigg\|\frac{1}{K} \sum_{k=1}^K \nabla F_k(\bm{w}_k(t,\zeta)) \bigg\|_2^2  \notag
\\&-  L^2 \frac{1}{K} \sum_{k=1}^K  \|( \bm{w}(t)-    \bm{w}_k(t,\zeta)) \|_2^2\bigg] \notag
= \frac{1}{2} \phi
\| \nabla F(\bm{w}(t)) \|_2^2 +
\frac{1}{2}  \sum_{\zeta=0}^{\phi-1}  \bigg\|\frac{1}{K} \sum_{k=1}^K \nabla F_k(\bm{w}_k(t,\zeta)) \bigg\|_2^2  \notag
\\&-  L^2  \frac{1}{2}  \sum_{\zeta=0}^{\phi-1} \frac{1}{K} \sum_{k=1}^K  \|( \bm{w}(t)-    \bm{w}_k(t,\zeta)) \|_2^2,
\end{align}
where $(a)$ is due to Assumption \ref{assumption3}, $(b)$ is due to  
$\bm{a}^T\bm{b}=\frac{1}{2}\|\bm{a}\|_2^2 + \frac{1}{2} \|\bm{b}\|_2^2 - \frac{1}{2}\|\bm{a}-\bm{b} \|_2^2$, 
and $(c)$ is due to Assumption \ref{assumption2}.

\noindent\textbf{Proof of Lemma 3.}
According to the definition of $\bm{\theta}(t)$, we have
\begin{align}
&\mathbb{E} \bigg[\|\bm{\theta}(t)  \|_2^2\bigg] \notag
= \mathbb{E} \bigg[\bigg\| \frac{1}{K}\sum_{k=1}^K \sum_{\zeta=0}^{\phi-1} \tilde{\bm{g}}_k(t,\zeta) \bigg\|_2^2\bigg] \notag
\mathop{=}^{(a)} \text{Var}\bigg(\frac{1}{K}\sum_{k=1}^K \sum_{\zeta=0}^{\phi-1} \tilde{\bm{g}}_k(t,\zeta) \bigg) 
+ \bigg\| \mathbb{E}  \bigg[\frac{1}{K}\sum_{k=1}^K \sum_{\zeta=0}^{\phi-1} \tilde{\bm{g}}_k(t,\zeta)\bigg] \bigg\|_2^2 \notag
\\& \mathop{\leq}^{(b)}\! \frac{1}{K^2}\! \sum_{k=1}^K \!\text{Var}\bigg(\! \sum_{\zeta=0}^{\phi-1} \tilde{\bm{g}}_k(t,\zeta) \!\bigg) \!
+\! \phi \sum_{\zeta=0}^{\phi-1} \bigg\| \! \frac{1}{K} \!\sum_{k=1}^K \! \nabla F_k(\bm{w}_k(t,\zeta)) \!\bigg\|_2^2, 
\end{align}
where  $(a)$ is due to $\mathbb{E}[\|\bm{x}\|^2] = \text{Var}[\bm{x}] + \|\mathbb{E}[\bm{x}]\|^2$
and $(b)$ is due to $\text{Var}(\sum_{j=1}^{n}\bm{x}_j)=\sum_{j=1}^n \text{Var}(\bm{x}_j)$ with independent  $\{ \bm{x}_j \}$.
With Assumption \ref{assumption3}, we have
$
	\text{Var}\bigg( \sum_{\zeta=0}^{\phi-1} \tilde{\bm{g}}_k(t,\zeta) \bigg) 
	\mathop{=}^{(a)} \sum_{\zeta=0}^{\phi-1} \text{Var}(  \tilde{\bm{g}}_k(t,\zeta) )
	\leq \phi \xi^2,
$
which yields \eqref{eq45}.

\noindent\textbf{Proof of Lemma 4.}
From the definition of $\bm{e}(t)$, we have
\begin{align}
&\mathbb{E}	[\| \bm{e}(t) \|_2^2]  \notag
= \mathbb{E}	\bigg[\bigg\| \frac{1}{K} \pi(t) \bigg(\hat{\bm{s}}(t)-\bm{s}(t)\bigg) \bigg\|_2^2\bigg]  \notag
\leq \frac{\Gamma}{K^2}\mathbb{E}\bigg[\bigg\|   \bigg(\hat{\bm{s}}(t)-\bm{s}(t)\bigg) \bigg\|_2^2\bigg]\notag
\\&\leq \frac{\Gamma(K+1)}{K^2} \bigg\{ \sum_{k=1}^K  \mathbb{E}\bigg[\bigg\|  \left( \frac{\sqrt{p_k(t)}|h_k(t)|}{\sqrt{\eta(t)}}\bm{I} - \bm{I} \right) \bm{s}_k(t) \bigg\|_2^2\bigg] \notag
+  \mathbb{E}\bigg[\bigg\| \frac{\bm{n}(t)}{\sqrt{\eta(t)}} 
\bigg\|_2^2\bigg] \bigg\} \notag
\\&=\frac{\Gamma(K+1)}{K^2} \bigg\{ \sum_{k=1}^K  N \bigg( \frac{\sqrt{p_k(t)}|h_k(t)|}{\sqrt{\eta(t)}} - 1 \bigg)^2 
+  \frac{N\sigma^2}{\eta(t)} \bigg\} 
=N \frac{\Gamma(K+1)}{K^2} \text{MSE}(t).
\end{align}

\bibliographystyle{IEEEtran}
\bibliography{ref}

\end{document}